\documentclass[prb,aps,twocolumn,superscriptaddress,showpacs]{revtex4}
\usepackage{graphicx,rotating,subfigure,amsmath,amsfonts,amssymb,delarray,pstricks}
\usepackage{array}
\usepackage{booktabs}
\usepackage{dcolumn}
\usepackage{units}
\usepackage{multirow}

\renewcommand{\vec}[1]{\boldsymbol #1}

\def\12{\frac{1}{2}}
\begin{document}
\bibliographystyle{apsrev}
%\nosight{*}

%\title{Physical properties of a dimerized ferromagnetic chain}
%% \title{Spin correlations in a dimerized ferromagnetic Heisenberg chain}
\title{The dimerized ferromagnetic Heisenberg chain}

\author{Alexander Herzog}
\affiliation{Department of Physics,
Technical University Kaiserslautern, D-67663 Kaiserslautern,
Germany}

\author{Peter Horsch}
\affiliation{ Max-Planck-Institut f\"ur Festk\"orperforschung,
            Heisenbergstrasse 1, D-70569 Stuttgart, Germany }

\author{Andrzej M. Ole\'s}
\affiliation{ Max-Planck-Institut f\"ur Festk\"orperforschung,
            Heisenbergstrasse 1, D-70569 Stuttgart, Germany }
\affiliation{ Marian Smoluchowski Institute of Physics, Jagellonian
            University, Reymonta 4, PL-30059 Krak\'ow, Poland }

\author{Jesko Sirker}
%% \email[]{Corresponding author: j.sirker@fkf.mpg.de}
\affiliation{Department of Physics, Technical University
Kaiserslautern, D-67663 Kaiserslautern, Germany}
\affiliation{Research Center OPTIMAS, Technical University
Kaiserslautern, D-67663 Kaiserslautern, Germany} \date{\today}

\begin{abstract}
%%PH
%%JS
Ferromagnetic, in contrast to antiferromagnetic, Heisenberg chains
can undergo a Spin-Peierls dimerization only at finite temperatures.
They show reentrant behavior as a function of temperature, which might play
%
%  For a ferromagnetic Heisenberg chain, a dimerization leads to a gain
%  in magnetic energy at finite temperatures. This mechanism might play 
%
a role for systems with small effective elastic constants as, for
example, monatomic chains on surfaces. We investigate the physical
properties of the dimerized ferromagnetic Heisenberg chain using a
modified spin-wave theory. % and the numerical density-matrix
% renormalization group applied to transfer matrices.  
We calculate 
%%PH
the exponentially decaying
spin and dimer correlation functions, analyze the
temperature dependence of the corresponding coherence lengths, the
susceptibility, as well as the static and dynamic spin structure
factor. By comparing with numerical data obtained by the
density-matrix renormalization group applied to transfer matrices,
we find that the modified spin wave theory yields excellent results
for all these quantities for a wide range of dimerizations and temperatures.
% We investigate the physical consequences of dimerization in a one-dimensional
% Heisenberg ferromagnet and show that spin
% correlations and their difference between even and odd bonds (i)
% decrease exponentially in the long range limit with the same correlation
% length, and (ii) their decrease is faster for increasing dimerization.
% Analytic results obtained using the modified spin-wave theory are in excellent
% agreement with the transfer matrix renormalization group.
% %Dimerization in
% %the ferromagnetic chain is responsible for two maxima in a
% %dynamical spin structure factor, while the static structure factor
% %is broadened when dimerization increases. Finally, we show that
% %the divergent behavior of the susceptibility in suppressed in a
% %dimerized chain.
% We show that increasing dimerization suppresses the
% divergent behavior of the susceptibility and is responsible for the
% broadening of the static structure factor which in addition exhibits a
% high-momentum peak for sufficiently strong dimerization and low temperature.
% Finally, we demonstrate that dimerization in the ferromagnetic chain is
% responsible for two maxima in a dynamical spin structure factor.
% %Finite temperature spin-Peierls instability in a ferromagnetic
% %spin chain. MSWT. Review of existing work. Review dynamic spin
% %structure factor of the uniform chain. New results for the
% %dimerized case. Thermodynamics
\end{abstract}


\maketitle

\section{Introduction}
\label{Intro}
One-dimensional (1D) spin systems play an important role in quantum
magnetism. A 
%%PH
basic
%paradigmatic 
model is the antiferromagnetic (AF) spin-$1/2$ Heisenberg chain which
is exactly solvable by Bethe ansatz \cite{Bethe} and has gapless
excitations (spinons). Many realizations of quasi 1D AF spin-$1/2$
Heisenberg chains are known today with Sr$_2$CuO$_3$ being one of the
best studied examples.\cite{MotoyamaUchida} An analysis in terms of a
spin-only model, however, might not always be applicable because of a
coupling to lattice degrees of freedom. In complete analogy to the
well-known Peierls effect\cite{Peierls} for 1D
conductors,\cite{SuShriefferHeeger,Hor81,Kiv82,Hir83,Bae85,Mal03} a
spin-Peierls transition can occur leading to a dimerization of the
spin exchange and a gapped excitation spectrum.\cite{CrossFisher} A
well known example is the spin-Peierls transition in
CuGeO$_3$.\cite{Has93}

In contrast, ferromagnetic (FM) spin chains are less frequent. An
example which has been analyzed in some detail is the organic system
(CH$_3$)$_4$NCuCl$_3$.\cite{LandeeWillett79} The magnetization and
susceptibility data of this system are well described by the
spin-$1/2$ FM Heisenberg chain. The thermodynamic properties of this
%%JS
fundamental 
model can be calculated by Bethe
ansatz\cite{TakahashiProgTheorPhys71,tak99} as well as by a modified
spin-wave theory (MSWT).\cite{MSWT} Further examples for realizations
of a 1D Heisenberg model with FM exchange are CuCl$_2$(DMSO) and
CuCl$_2$(TMSO),\cite{SwankLandee79} where DMSO (TMSO) stands for
dimethylsulfoxide (tetramethylsulfoxide), (C$_6$H$_11$NH$_3$)CuCl$_3$
and (C$_6$H$_11$NH$_3$)CuBr$_3$,\cite{KopingaTinus81} as well as
2-benzimidazolyl nitronyl nitroxide (2-BIMNN).\cite{SuganoBlundell}

Another fascinating system are 1D arrays of Co chains which
self-assemble on Pt substrates. The magnetic exchange in these
nanostructured monatomic chains has been shown to be predominantly
FM.\cite{Gambardella,VindingiRettori2006}
%  Albeit it is not a priori clear that the
% Heisenberg Hamiltonian captures the physics of a metallic
% ferromagnetic chain, this model seems to describe the magnetism of the
% system well.\cite{VindingiRettori2006}
% Indeed for low temperature it is possible to identify the free energy of an
% itinerant ferromagnet with the free energy of a classical Heisenberg Hamiltonian.
% \cite{Cyrot, PrangeKorenmann78, Mathon82}
The specific spin texture in such systems, however, might be
complicated\cite{WiesendangerRMP} due to the reduced symmetry allowing
for exchange interactions different from a pure Heisenberg exchange.
Since the atoms can be easily moved along the surface, the effective
elastic constants of these monatomic chains are expected to be small.
This has lead us, \textit{inter alia}, to the question if a coupling
to lattice degrees of freedom can induce a dimerization of FM exchange
interactions similar to the AF case.

In a recent work,\cite{SPPRL} we have shown that such a spin-Peierls effect is
indeed possible for the FM Heisenberg chain---at least at the level of the adiabatic
approximation---but has to be activated by thermal fluctuations. 
% This
% is due to a gain in magnetic energy $\sim T^{3/2}\delta^2$ accompanied
% by the dimerization which under
% given circumstances can outweigh the elastic energy cost of the lattice
% deformation. 
Furthermore, we have argued that this mechanism seems to explain the
observed dimerization of the FM exchange in the $C$-phase of the
perovskite YVO$_3$,\cite{PhysRevLett.91.257202,Hor03} with the
dimerization being caused by a coupling to orbital rather than lattice
degrees of freedom.

The aim of this paper is to study the physical properties of a
dimerized FM spin chain independent of the mechanism which causes the
dimerization.  As an analytical method to calculate the thermodynamic
properties of the dimerized ferromagnetic Heisenberg chain we will use
a MSWT and  compare with
%%PH
numerical data obtained by the transfer matrix renormalization group
(TMRG).
%%JS: Mir gefiel die vorherige Version besser
As a central result of our paper we show that the MSWT yields excellent 
results for a wide range of dimerization parameters and temperatures, i.e.,
when we compare with TMRG data. In particular, we find that at finite
temperatures $T$ both the spin and the dimer correlation function decay
exponentially with the same correlation length  (we set $\hbar=k_\textnormal{B}=1$)
\begin{equation}\label{Xi}
	\xi_{\delta} =c(T,\delta) JS^2/T,
\end{equation}
where $J<0$ is the FM exchange constant in the Heisenberg Hamiltonian and $S$ is the spin quantum number. The lattice spacing has been set to unity.
A detailed discussion of the function $
c(T,\delta)$ which captures the effects of the dimerization $\delta$
as well as the higher order temperature dependence of the correlation
length is given for $S=1/2$ and $S=1$.  Moreover, we also present
explicit analytical  expressions for the doping and
temperature dependence of the asymptotic behavior of correlation functions
and the spin susceptibility.

Then we address the
question how the spin-Peierls symmetry breaking manifests itself in
the spin-structure factor given that the correlation functions decay
exponentially. To detect this symmetry breaking we show that it is
useful to study an off-diagonal spin-structure factor.
Three approaches are used to calculate the static structure factors:
(i) via the dynamic spin structure factors, and via
the equal-time spin correlation functions as obtained by (ii) MSWT
and (iii) TMRG.

The paper is organized as follows: In Sec.~\ref{SpinPeierls} we
introduce and motivate the model.  Furthermore, we discuss the set-up
of a MSWT which will be used throughout the paper to obtain analytical
results for various thermodynamic quantities. In Sec.~\ref{sec:ss}
spin correlation functions for the uniform and the dimerized 1D chain
are discussed. Here we derive analytic expressions for the correlation
functions both in the short and the long-range limit, calculate the
magnetic susceptibility and compare with data obtained from numerical
TMRG calculations. In Sec.~\ref{sec:dyn} and Sec.~\ref{sec:sta} we
discuss the dynamic and the static spin structure factor,
respectively. In Sec.~\ref{sec:sum} we present a short discussion and
a summary of our results.

\section{Spin-Peierls phase in 1D ferromagnets}
\label{SpinPeierls}

The 
%%PH
Spin-Peierls
Hamiltonian we want to investigate reads
\begin{equation}\label{H}
H=-J\sum_{j=1}^N\{1+(-1)^j\delta\}\vec S_j\cdot\vec S_{j+1},
\end{equation}
where $J>0$ is the nearest neighbor exchange interaction among the spin S
operators $\vec S_j$ and $\vec S_{j+1}$ on the sites $j$ and $j+1$,
respectively, and $\delta\in[0,1]$ is the dimerization parameter.
%$\delta$ is the dimerization parameter,
%%PH
Consistent with dimerization we consider an even number $N$ of spins and 
assume periodic boundary conditions.
%
%%JS
In Ref.~\onlinecite{SPPRL} the Hamiltonian, Eq.~(\ref{H}), was
investigated by TMRG. 
This method is particularly suited to study the
properties of 1D quantum systems directly in the \textit{thermodynamic
 limit}.  Details about this approach are given in
Refs.~\onlinecite{PeschelWang,SirkerKluemperPRB66,SirkerKluemperEPL,GlockeKluemper}.
In particular the gain in the free energy $\delta
f\sim-T^{3/2}\delta^2$ stemming from a dimerization was compared in an
adiabatic approximation to the cost in elastic energy $\sim K\delta^2$
caused by the lattice distortion.\cite{energy} A dimerized phase was
found to be stable in a regime of finite temperatures for a
sufficiently small elastic constant
%$K/|J|\leq K_c/|J|\simeq 0.118$.
$K\leq K_c$, with $K_c/J\simeq 0.118$. 
% This result can be understood within the
% framework of the MSWT which predicts that the gain in free energy
% $\delta f\sim-T^{3/2}\delta^2$ due to the
% dimerization is of the same order in $\delta$ as the elastic energy cost.\cite{energy}
% Moreover, because of the temperature dependence of the magnetic energy gain
% this transition has to be thermally activated.

In order to apply the MSWT, we introduce
boson operators $\{a_{m}^{\dagger},a_{m}^{\ }\}$ on site $m$ via a Dyson-Maleev transformation
\begin{equation}\label{dama}
	\begin{aligned}
		&S_{m}^-=a_{m}^{\dagger}\,,\nonumber\\
		&S_{m}^+=(2S-a_{m}^{\dagger}a_{m}^{})a_{m}^{}\,,\nonumber\\
		&S_{m}^z=  S-a_{m}^{\dagger}a_{m}^{}.
	\end{aligned}
\end{equation}
Using the Dyson-Maleev transformation for the Hamiltonian,
Eq.~(\ref{H}), and expanding to bilinear order in the boson operators,
$H$ is easily diagonalized in terms of new boson operators
$\{\alpha_k^{\dagger},\ \alpha_k^{\ };\ \beta_k^{\dagger},\ \beta_k^{\
}\}$ obtained by subsequent Fourier and Bogoliubov transformation. The
introduction of two sets of bosonic operators is required here since
the unit cell of a dimerized FM chain is doubled. We find
\begin{equation}\label{HMSWT2}
H_S=E_S^0+\sum_k\left\{\omega_k^-\alpha_k^\dagger\alpha_k^{\ }
+\omega_k^+\beta_k^\dagger\beta_k^{\ }\right\}\,.
\end{equation}
\begin{figure}[!t]
		\includegraphics[width=.95\linewidth]{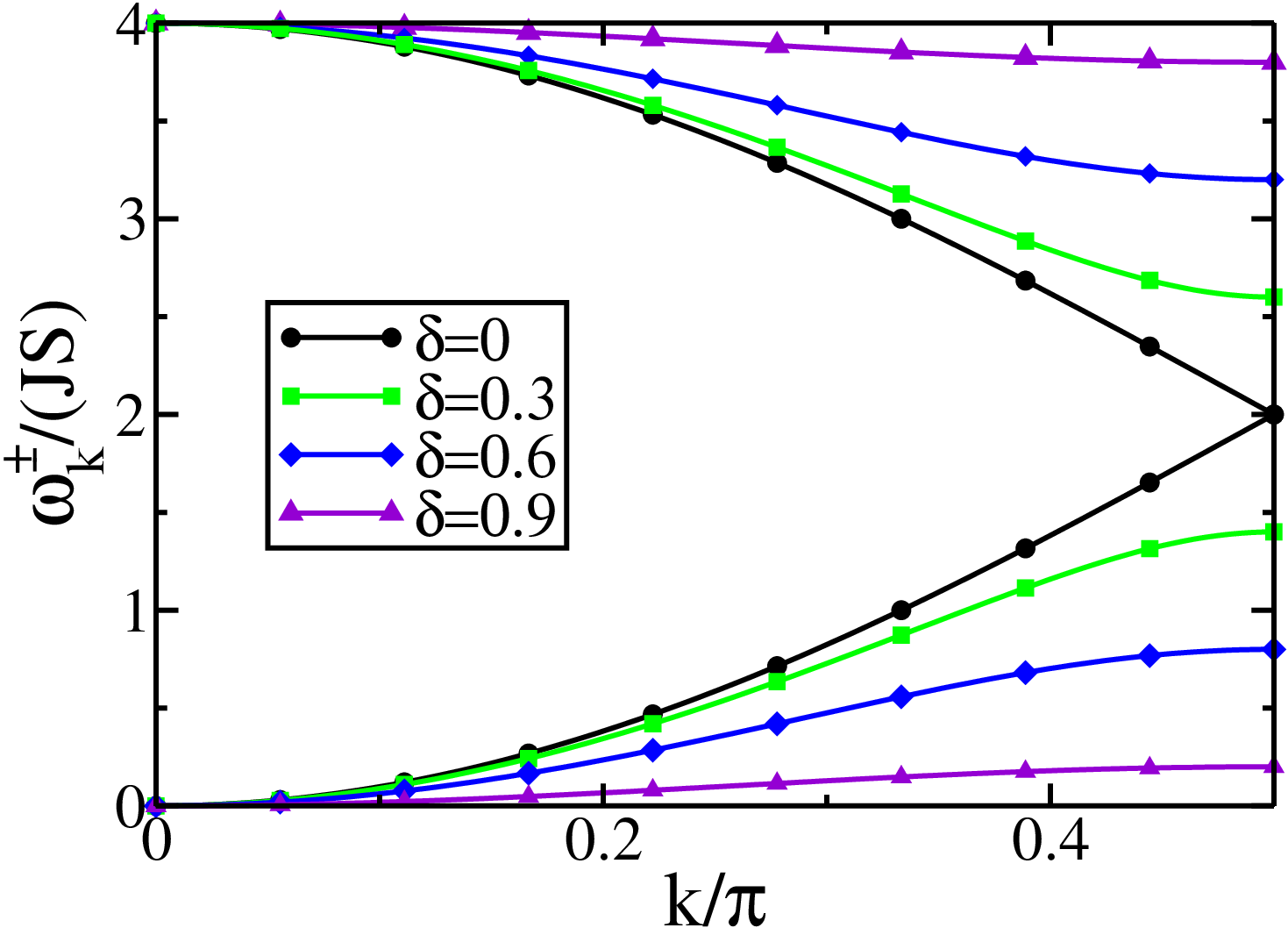}
  \caption{(Color online) Magnon dispersions obtained
  from MSWT for various dimerizations.
  The lower branches correspond to $\omega_k^-$ while
  the upper branches are given by
  $\omega_k^+$.}\label{Disp}
\end{figure}
\begin{figure*}[t]
\includegraphics[width=.9\linewidth]{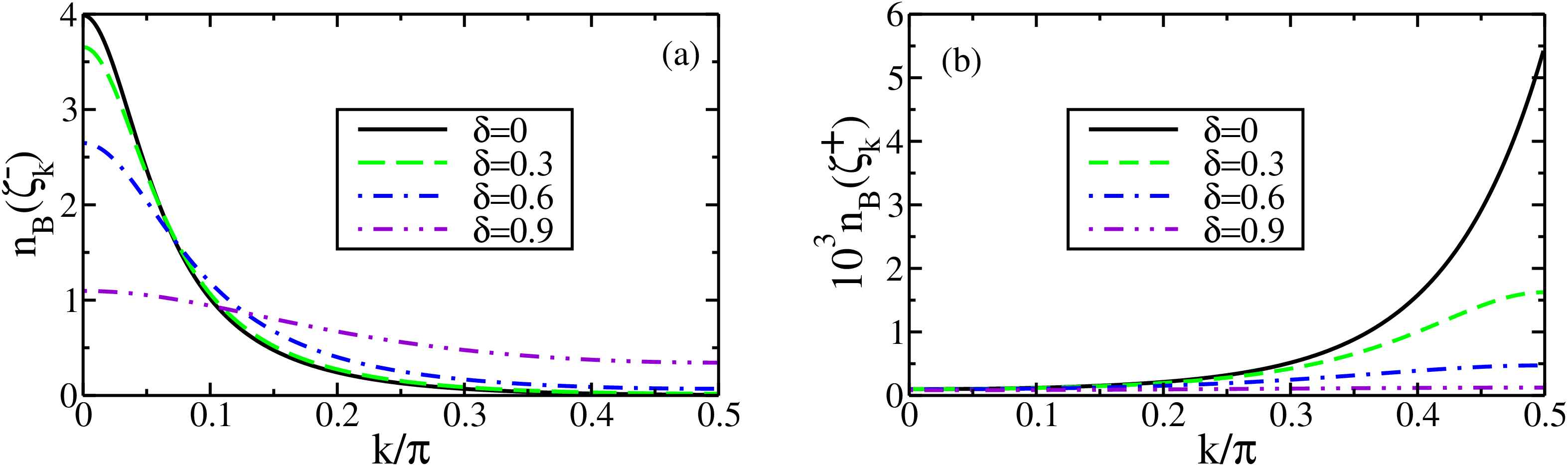}
\caption{(Color online) Occupation numbers for the (a) lower and (b)
  upper branch for $S=1$ and $T=0.5J$ and various dimerizations as
  indicated.}\label{Oc}
\end{figure*}%       in one-dimensional ferromagnets}
Here $E_S^0=-JNS^2$ is the ground state energy at zero temperature
($T=0$), which is independent of the dimerization $\delta$, $k$ is the
1D momentum, and $\omega_k^\pm$ stands for the dispersion of the two
magnon branches given by
\begin{equation}\label{omegapm}
\omega_k^\pm=2JS\left(1\pm\sqrt{\cos^2k+\delta^2\sin^2k}\,\right)\,.
\end{equation}
From this expression it becomes clear that a finite dimerization,
$\delta>0$, induces a splitting of the spin-wave dispersion into an
accoustic and an optical branch at the boundaries $k=\pm\pi/2$ of the
Brillouin zone. As a consequence, the two branches flatten with
increasing dimerization. This is shown in Fig.~\ref{Disp} where the
dispersions for various values of $\delta$ are shown.

To fulfill the Mermin-Wagner theorem of vanishing magnetization at
finite temperature $T$, usual spin-wave theory has to be modified by a
Lagrange multiplier serving as a chemical potential $\mu_\delta(T)$ that
sets the finite temperature magnetization to zero.\cite{MSWT,MSWTlong}
This results in the constraint %local constraint at
%site $m$,
%\begin{equation}\label{MSWTlocal}
%\langle a_{m,l}^{\dagger} a_{m,l}^{\ } \rangle=S,
%\end{equation}
%
%which implies the global constraint in momentum space,
%
\begin{equation}\label{MSWTconstraint}
S=\frac{1}{N}\sum_k\{n_\textnormal{B}(\zeta_k^-)
+n_\textnormal{B}(\zeta_k^+)\}\,.
\end{equation}
%
% The constraint Eq.~(\ref{MSWTconstraint}) determines $\mu_\delta$ at
% any finite temperature $T$.
Here
\begin{equation}\label{bose}
n_\textnormal{B}(x)=\frac{1}{\exp(\beta x)-1}
\end{equation}
is the Bose distribution function, $\beta=1/T$, and
\begin{equation}\label{redu}
\zeta_k^\pm=\omega_k^\pm-\mu_{\delta}(T)
\end{equation}
is the reduced magnon dispersion. The chemical potential $\mu_\delta(T)$
will be negative and vanishes as $T\rightarrow 0$.

Results obtained from this procedure are in excellent agreement with
the Bethe ansatz results for the $S=1/2$ uniform
ferromagnet\cite{MSWT} ($\delta=0$) 
%%JS
% as well as with data
% obtained from the TMRG for arbitrary $S$ 
% both for the uniform as well
% as for the dimerized chain\cite{ICM} 
%
if $t_\delta/S\leq1$, where we
have defined a reduced temperature
\begin{equation}\label{td}
	t_{\delta}\equiv\frac{T}{JS(1-\delta^2)}.
\end{equation}
%%JS
The occupation numbers for the lower and upper branch
$n_\text{B}(\zeta_k^-)$ and $n_\text{B}(\zeta_k^+)$ are shown in
Fig.~\ref{Oc} for $T/J=0.5$ and $S=1$. 
As expected, the occupation for
the lower branch shows a maximum at $k=0$ whereas the occupation for
the upper branch has a maximum at $k=\pi/2$. With increasing
dimerization these peaks become less pronounced due to the flattening
of the dispersions, see Fig.~\ref{Disp}.
%Moreover at sufficiently low temperatures, due to the fact that the optical
%magnon branch is higher in energy than the accoustical one the occupation of
%the upper branch is much lower than the occupation for the lower branch.
%

%\section{Static spin-structure factor and thermodynamics}
%\label{TD}
\section{Correlation functions}
\label{sec:ss}

%\subsection{Uniform ferromagnetic chain} \label{sec:uni}%
To obtain the spin correlation function (SCF) for an arbitrary
distance $|i-j|$ it is crucial to include also the quartic terms in the bosonic
operators stemming from the Dyson-Maleev transformation of the
scalar product ${\bf S}_i\cdot{\bf S}_j$.
%\begin{eqnarray}
%\label{SiSj} {\bf S}_i{\bf
%S}_j&=&S^2-S(a_i^{\dagger}-a_j^{\dagger})(a_i-a_j)
%\nonumber \\
%&-&\frac{1}{4}\Big\{a_i^{\dagger}a_j^{\dagger}(a_i-a_j)^2
%   + (a_i^{\dagger}-a_j^{\dagger})^2a_ia_j \Big\},
%\end{eqnarray}
This leads to 
%
%\begin{equation}
%\label{SiSj_A} \langle{\bf S}_i\cdot{\bf S}_j\rangle= \langle a_{i}^\dagger a_{j}^{\ }\rangle\langle a^{\ }_{i}a_{j}^\dagger\rangle.
%\end{equation}

\begin{equation}
\label{SiSj_A}
\langle{\bf S}_i{\bf S}_j\rangle= \left\{
\begin{array}{rl}
&S(S+1) \mbox{ for } i=j\\ 
&\frac{1}{4}\Big(
\langle a_i^{\dagger} a_j\rangle + \langle a_j^{\dagger} a_i\rangle\Big)^2
\mbox{ for } i\ne j
\end{array}\right. .
\end{equation}
For $T>0$ the correlation function 
approaches zero in the limit $|i-j|\rightarrow \infty$ as
$\langle a_i^{\dagger} a_j\rangle$ vanishes in this case. 
% Inspection shows that the neglect of
% the $O(S^0)$ terms in the expansion of $\langle{\bf S}_i{\bf S}_j\rangle$
% would yield a wrong asymptotic value 
% $\langle{\bf S}_i{\bf S}_j\rangle \rightarrow -S^2$ instead of zero.

\begin{figure}[t]
	\includegraphics*[width=.85\linewidth]{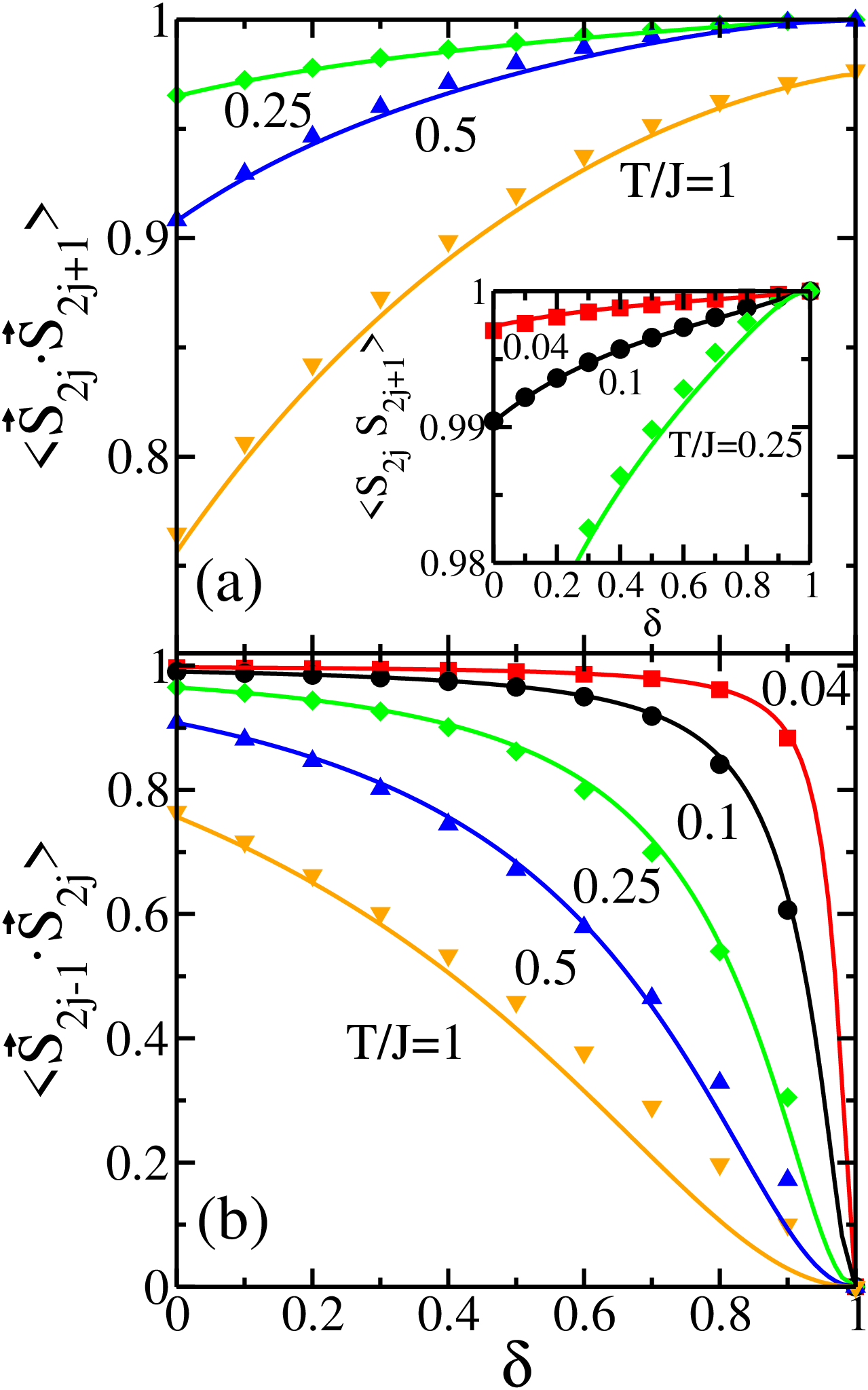}
	\caption{(Color online) Nearest neighbor spin correlation
        functions on (a) the strong bonds and (b) the weak bonds as
        a function of dimerization for various temperatures 
%%JS
and $S=1$. 
The solid lines show the MSWT results. For comparison, TMRG data are
shown as symbols. The inset shows data for low
temperatures.}\label{NNSSS1}
\end{figure}
%
%From this expression it is obvious that the SCF approaches zero for $|i-j|\rightarrow \infty$ as
%$\langle a_{i,l}^{\dagger}a_{j,l'}\rangle$ vanishes in
%this limit. We emphasize that the neglect of the
%$O(S^0)$ terms in the expansion of $\langle{\bf S}_i{\bf
%S}_j\rangle$ would yield instead an incorrect
%asymptotic behavior, with $\langle{\bf S}_i{\bf
%S}_j\rangle\rightarrow -S^2$.
\begin{figure*}[t]
	\begin{minipage}{.47\linewidth}
		\includegraphics*[width=8cm]{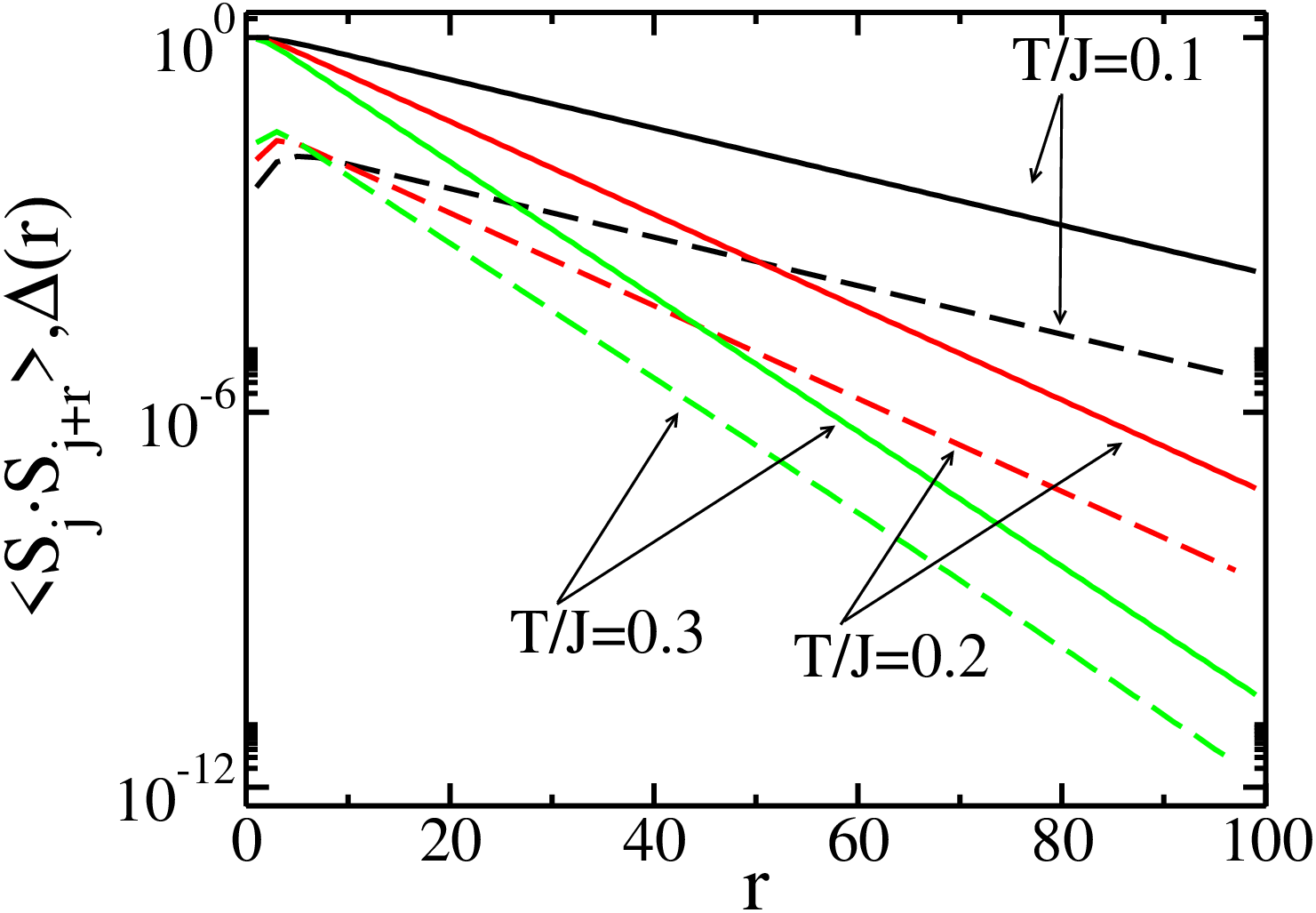}
		\caption{(Color online) Spin correlation function
                (solid lines) and alternation correlation function
                $\Delta(r)$ (dashed lines) for $S=1$ and
                $\delta=0.1$ and various temperatures as a function
                of distance $r$ obtained by MSWT.  For a given
                temperature, the exponential decay of both
                quantities is determined by the same correlation
                length.}\label{SSDELVT}
	\end{minipage}
	\hspace{.01\linewidth}
	\begin{minipage}{.47\linewidth}
		\includegraphics*[width=8cm]{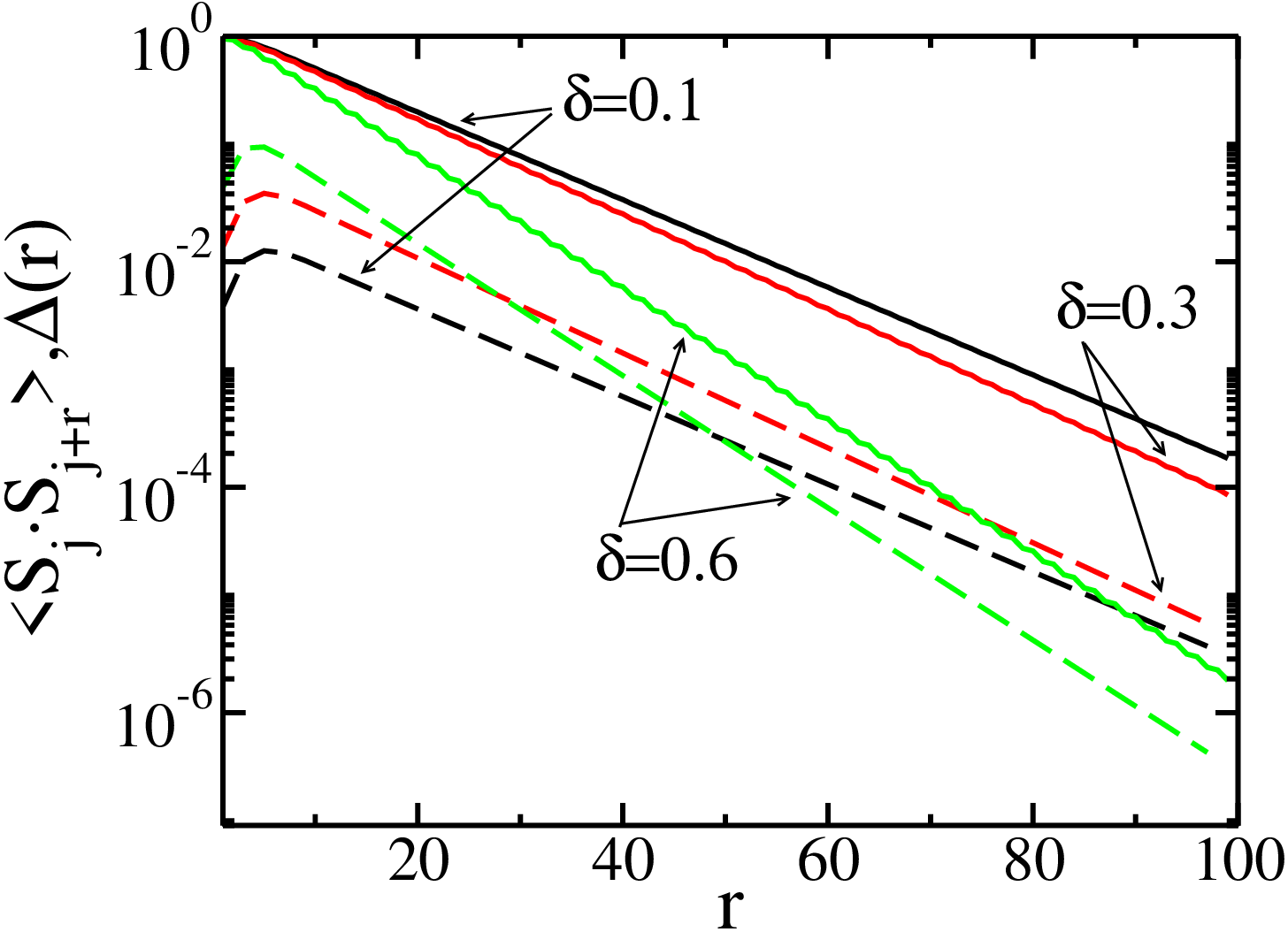}
		\caption{(Color online) Spin correlation function
                (solid lines) and alternation correlation function
                (dashed lines) obtained by MSWT for $S=1$ and
                $T/J=0.1$ and various dimerizations as
                indicated.}\label{SSDEL}
	\end{minipage}
\end{figure*}
%The static CF are obtained by taking the limit $\tau\rightarrow0$
%in the respective expression contained in Eq. (\ref{Gtau}).
%From this

For the dimerized chain we obtain from Eq.~(\ref{SiSj_A})
%It follows that $\left<(\vec S_j)^2\right>=S(S+1)$ and
\begin{equation}\label{MSWTCFo}
\left<\vec{S}_j\cdot\vec{S}_{j+r}\right>=\left(\frac{1}{N}\sum_k
\left[n_\textnormal{B}(\zeta^-_k)+n_\textnormal{B}(\zeta_k^+)\right]
\cos(kr)\right)^2,
\end{equation}
if $r\ne0$ is even, and
\begin{equation}\label{MSWTCFe}
  \left<\vec{S}_j\cdot\vec{S}_{j+r}\right>=
\left(\frac{1}{N}\sum_k\left[n_\textnormal{B}(\zeta^-_k)
-n_\textnormal{B}(\zeta_k^+)\right]f_k(j,r,\delta)\right)^2,
\end{equation}
if $r$ is odd. Here we have defined
\begin{equation}\label{fjr}
f_k(j,r,\delta)\equiv\frac{\cos(kr)\cos k+(-1)^j\delta\sin(kr)\sin
k} {\sqrt{\cos^2k+\delta^2\sin^2k}}.
\end{equation}
Note that the summation in Eqs.~(\ref{MSWTCFo}) and (\ref{MSWTCFe})
runs over the reduced Brillouin zone which is the reason for the minus
sign in Eq.~(\ref{MSWTCFe}). From these expressions we obtain the
nearest neighbor SCF shown in Fig.~\ref{NNSSS1}.  One finds that the
correlations on the strong bonds are enhanced while those on the weak
bonds are reduced when the dimerization is increased. The dimerization
strongly affects the SCF on the weak bonds whereas the strong bonds
are far less affected.  We find excellent agreement between the
results obtained by MSWT and TMRG data up to $T/J=1$ and even the
value at full dimerization ($\delta=1$) is captured correctly.
However, MSWT suggests a quadratic behavior in
$\epsilon\equiv1-\delta$ on the weak bonds for $\epsilon\ll1$ whereas
a perturbative treatment starting from decoupled dimers shows a linear
behavior as confirmed by the TMRG data. Finally the steep decrease of
the nearest neighbor SCF on the weak bonds for low temperatures and
strong dimerization indicates a non-analyticity for $T=0$ at
$\delta=1$.
%%PH
Thus the ground state of the dimerized chain is always
a uniform ferromagnetic state except for $\delta=1$ where it consists of
decoupled dimers.

Next we shall explore the long-distance behavior of the spin-correlation
function $\left<\vec{S}_j\cdot\vec{S}_{j+r}\right>$
and its variation with dimerization $\delta$ and temperature. In addition
we define an alternation correlation function (ACF) for odd $r$ as
\begin{equation}\label{Delta}
  \Delta(r)\equiv|\left<\vec{S}_j\cdot\vec{S}_{j+r}\right>-
  \left<\vec{S}_j\cdot\vec{S}_{j-r}\right>|.
\end{equation}
This expression describes the alternation of the SCF between the weak
and the strong bonds and is thus a measure for the dimerization of the
system. 
%%PH
We exclude even $r$ in  Eq.~(\ref{Delta}), as in this case
$\Delta(r)=0$ by symmetry.
In Fig.~\ref{SSDELVT} the SCF and the ACF obtained from MSWT
for $S=1$ and $\delta=0.1$ are shown for different temperatures and in
Fig.~\ref{SSDEL} for $S=1$ and $T/J=0.1$ and different dimerizations.
The asymptotic behavior of the correlation functions is governed in
both cases by an exponential decay $\sim \text{e}^{-|r|/\xi_\delta}$
% $$
% 	\left<\vec S_j\cdot\vec S_{j+r}\right>\sim\text{e}^{-|r|/\xi_\delta}
% $$
% and
% $$
% 	\Delta(r)\sim\text{e}^{-|r|/\xi_\delta}
% $$
with the {\it same} correlation length $\xi_\delta$.
As expected, increasing dimerization $\delta$ reduces the correlation
length. For large dimerization the alternation of the
exchange integrals
leads to a staircase-like behavior of the SCF $\left<\vec{S}_j\cdot\vec{S}_{j+r}\right>$ (see
Fig.~\ref{SSDEL} for $\delta=0.6$).

\begin{figure*}[t]
\begin{minipage}{.48\linewidth}
\includegraphics[width=8cm]{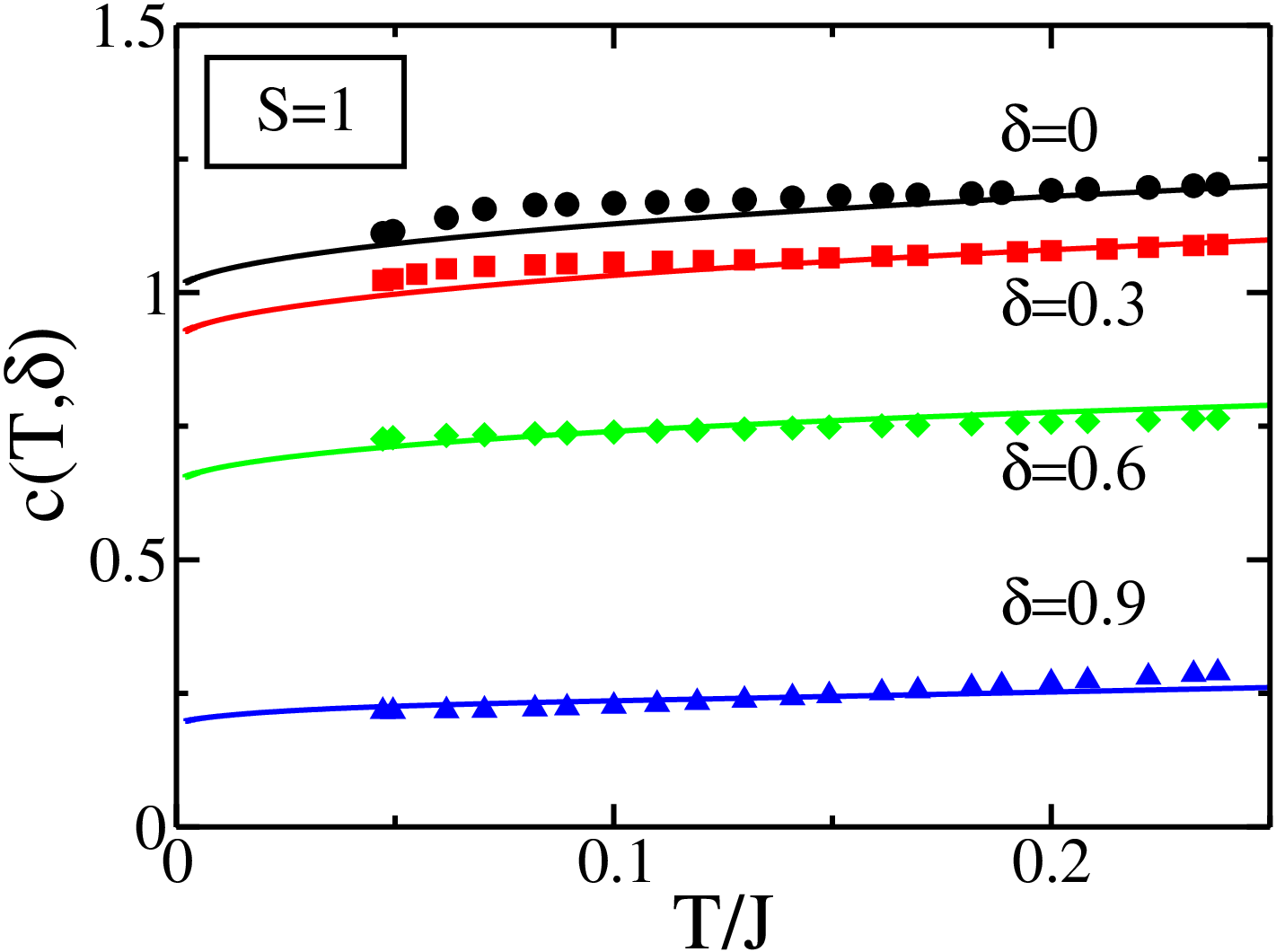}
\caption{(Color online) Comparison of the function $c(T,\delta)$ as
given by Eq.~(\ref{can}) (lines) and TMRG data (symbols) for $S=1$
and various $\delta$ as indicated.}\label{C1}
\end{minipage}\hspace{.01\linewidth}
\begin{minipage}{.48\linewidth}
	\includegraphics[width=.95\linewidth]{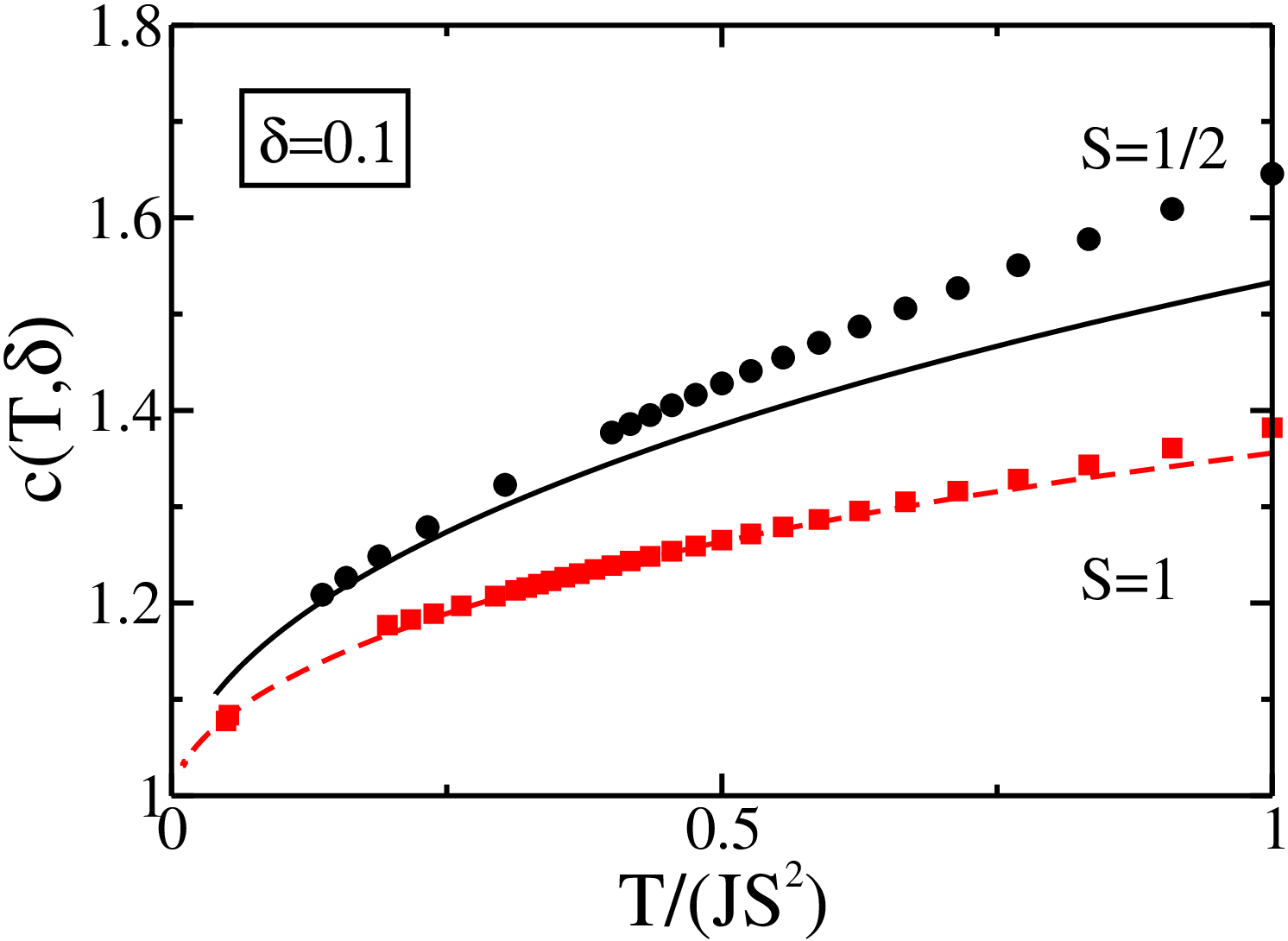}
	\caption{(Color online) Comparison of $c(T,\delta)$ for
        $S=1/2$ and $S=1$ as a function of temperature at
        $\delta=0.1$. The MSWT results (lines) are compared with
        TMRG data (circles and squares).}
	\label{ccomparison}
\end{minipage}
\end{figure*}

The function $c(T,\delta)$ (see Eq.~(\ref{Xi})) describes the behavior
of the correlation length as a function of dimerization and also gives
corrections to the leading order $1/T$ temperature dependence. For a
uniform 1D ferromagnet to lowest order in $T$ we have
$c(T,\delta=0)=1$.\cite{MSWTlong} In order to go beyond this limit we
have to investigate in detail the behavior of the SCF at large
distances. For sufficiently low temperatures we may neglect
contributions stemming from $n_\textnormal{B}(\zeta_k^+)$ in
Eqs.~(\ref{MSWTconstraint}), (\ref{MSWTCFo}), and (\ref{MSWTCFe}).
Expanding $\omega_k^-$ and $f_k(j,r,\delta)$ to lowest non-vanishing
order and using a saddle point integration in Eqs.~(\ref{MSWTCFo}) and
(\ref{MSWTCFe}) we obtain in a large-$r$ expansion ($r\gg1$)
\begin{equation}\label{SSdlr}
\begin{aligned}
\left<\mathbf{S}_j\cdot\mathbf{S}_{j+r}\right>\approx&
\frac{t_{\delta}}{4}\left(\frac{1}{\sqrt{v_{\delta}}}
+\text{sgn}(r)\frac{(-1)^j(1-(-1)^r)}{2}\delta\sqrt{t_{\delta}}\right)^2\\
&\times\textnormal{e}^{-|r|/\xi_{\delta}},
\end{aligned}
\end{equation}
with the correlation length 
\begin{equation}\label{CL2}
	\xi_{\delta}=\frac{1}{2\sqrt{v_{\delta}t_{\delta}}}.
\end{equation}
Here $v_\delta=-\mu_\delta/T$ is the negative reduced
chemical potential reading
\begin{equation}\label{van}
v_\delta=\frac{t_\delta}{4S^2}\left(1+\alpha\sqrt{t_\delta}
+\frac{3}{4}\alpha^2t_\delta+\frac{\alpha^{3}}{2}t_\delta^{3/2}+\dots\right),
\end{equation}
with $\alpha\equiv\zeta(\frac{1}{2})/(S\sqrt{\pi})$ where $\zeta(z)$
is Riemann's zeta function. Eq.~(\ref{SSdlr}) explains why the SCF and
the ACF both decay with the same correlation length. We note that for
$\delta=0$ this expression agrees with the one obtained for the
uniform case.\cite{BortzSirker}
\begin{table}[!t]
\begin{ruledtabular}
\begin{tabular}{rl}
&
\begin{tabular}{lllll}
	$\delta$\quad\ & $a_1$\qquad\quad\  & $a_2$\qquad\quad\  & $a_3$\quad\qquad\ & $a_4$\qquad
\end{tabular}
\\\hline
\Large $S=\frac{1}{2}$ & \normalsize
\begin{tabular}{lllll}
	 
	 0\ \  & 1.00000\ \  & 1.16519\ \  & 4.60822 \ &9.66505\\
	 0.3\ \  & 0.91000\ \ & 1.11152\ \  & 5.06398 \ &11.1338\\
	 0.6\ \  & 0.64000\ \  & 0.93216\ \  & 7.20035 \ &18.8771\\
	 0.9\ \  & 0.19000\ \  & 0.50710\ \  & 24.2538 \ &116.701\\
\end{tabular}\\\hline
\Large $S=1$ & \normalsize
\begin{tabular}{lllll}
	 0\ \  & 1.00000\ \  & 0.41195\ \  & 0.07200\ \ &0.05339\\
	 0.3\ \  & 0.91000\ \ & 0.39298\ \  & 0.07912\ \ &0.06151\\
	 0.6\ \  & 0.64000\ \  & 0.32956\ \  & 0.11251\ \ &0.10428\\
	 0.9\ \  & 0.19000\ \  & 0.17956\ \  & 0.37897\ \ &0.64469\\
\end{tabular}
\end{tabular}
\end{ruledtabular}
\caption{Coefficients of $c(T,\delta)$ given in Eq.~(\ref{cT}) as extracted from Eq.~(\ref{can}) for $S=1/2$ and $S=1$. The coefficient $a_1$ is given by ($1-\delta^2$).}\label{Tab1}
\end{table}
\begin{figure}[t]
	%\begin{minipage}{.49\linewidth}
	\includegraphics[width=8cm]{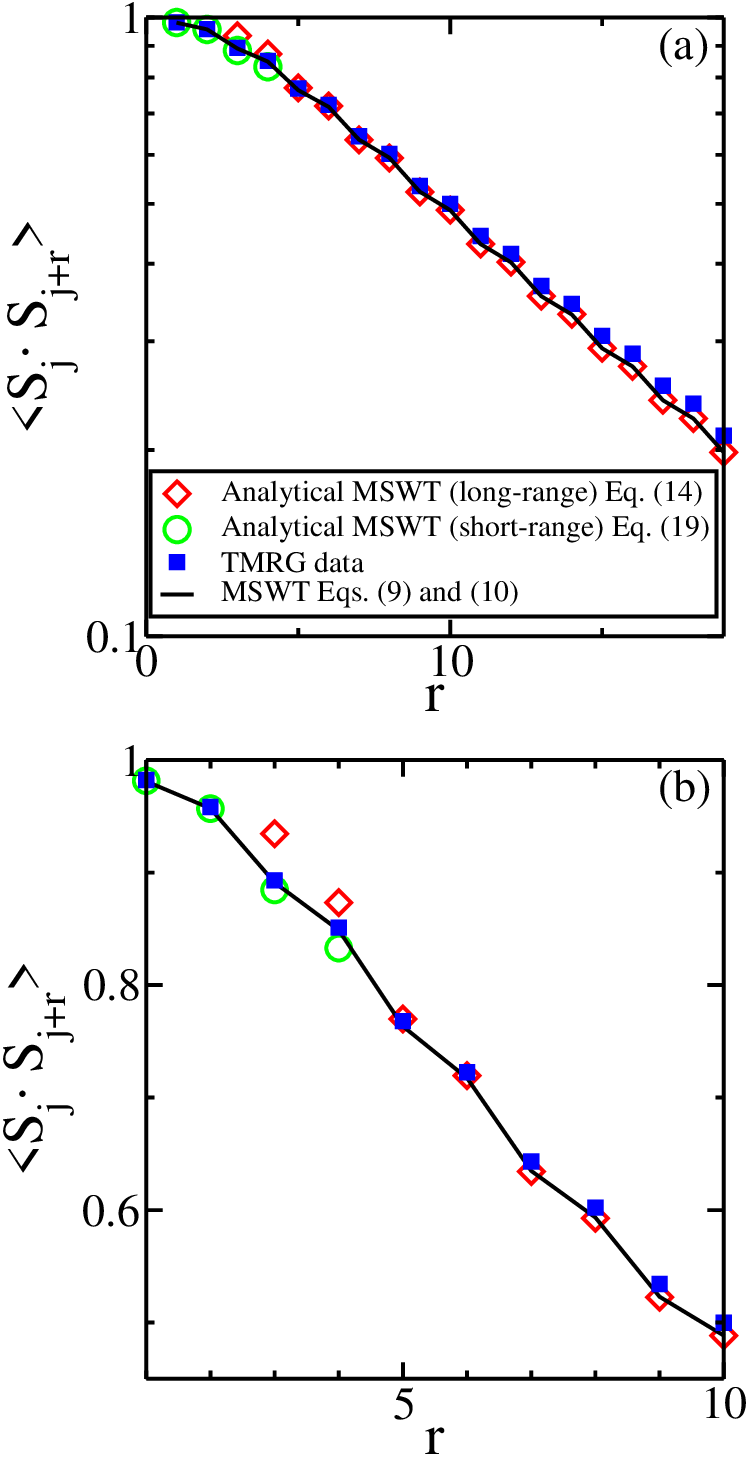}
	\caption{(Color online) (a) Spin correlation function
        $\left<\vec S_j\cdot\vec S_{j+r}\right>$ on a logarithmic
        scale for $S=1$, $T/J=0.1$ and $\delta=0.3$ obtained by TMRG compared
        with the analytical results from the asymptotic expansion,
        Eq.~(\ref{SSdlr}), and the short-range expression, Eq.
        (\ref{SSsr}), up to $r=4$. The solid line connects data
        obtained from a numerical solution of the MSWT,
        Eqs.~(\ref{MSWTCFo}) and (\ref{MSWTCFe}), for integer $r$.
        (b) Same as above on a linear scale.}\label{LogCF}
	%\end{minipage}
\end{figure}
\begin{figure}
	%\begin{minipage}{.48\linewidth}
		\ \vspace*{.2cm} \\\includegraphics*[width=.95\linewidth]{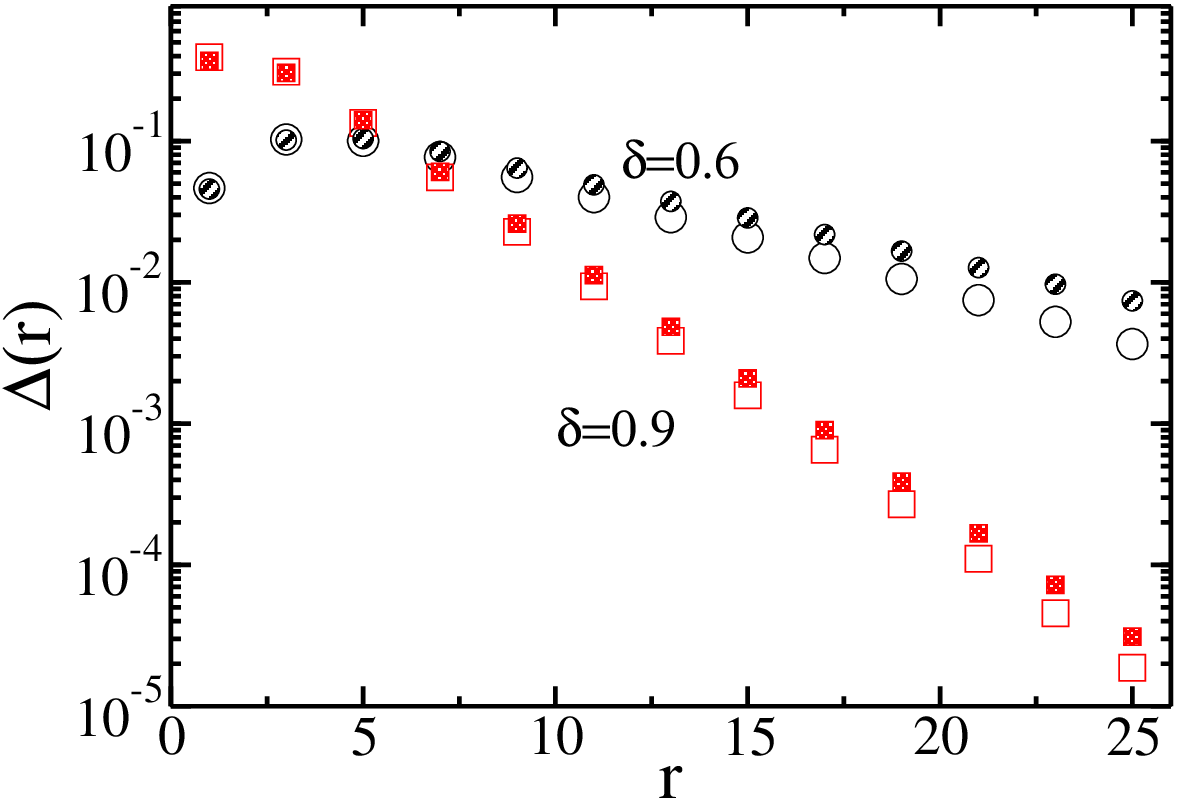}
		\caption{(Color online) Difference between the spin
                correlations on weak and strong bonds, see
                Eq.~(\ref{Delta}), for $S=1$, $T/J=0.1$, and two
                representative dimerization parameters. The filled
                symbols are data obtained from TMRG whereas the open
                symbols show results obtained by
                MSWT.\\\\}\label{lndell}
	%\end{minipage}
\end{figure}

Inserting Eq.~(\ref{van}) into Eq.~(\ref{CL2}) and expanding for 
$t_\delta\ll1$ we obtain the analytical expression % for  $c(T,\delta)$ reading
\begin{equation}\label{can}	
	c(T,\delta)\approx(1-\delta^2)
\left(1-\frac{\alpha}{2}\sqrt{t_\delta}+\frac{5\alpha^4}{32}t_\delta^2-\frac{9\alpha^5}{64}t_\delta^{5/2}\right).
\end{equation}
In Fig.~\ref{C1}, $c(T,\delta)$ for $S=1$ and various dimerizations is
shown in comparison to numerical data from TMRG. We find excellent
agreement between both methods.
%In Tab.~\ref{Tab1} these analytical results are compared to a fit obtained from the TMRG data obtained from
%\begin{equation}\label{cfit}
%	c(\delta,T)=a_0+a_1T^{1/2}+a_2T+a_3T^{3/2}+a_4T^2.
%\end{equation}
In Fig.~\ref{ccomparison}, $c(T,\delta)$ is shown for $S=1/2$ and
$S=1$ as a function of temperature for $\delta=0.1$. 
% For $S=1/2$ the
% agreement between the analytical result and the TMRG data is only good
% at lower temperatures. 
Note that $c(T,\delta)$ is much steeper for $S=1/2$, see
Eq.~(\ref{can}). In Tab.~\ref{Tab1} the coefficients $a_1$, $a_2$,
$a_3$, and $a_4$ of the series
\begin{equation}\label{cT}
	c(T,\delta)\approx a_1+a_2\sqrt{\frac{T}{J}}+a_3\left(\frac{T}{J}\right)^2+a_4\left(\frac{T}{J}\right)^{5/2}
\end{equation}
are shown as extracted from Eq.~(\ref{can}). We remark that the higher
order corrections $a_3$ and $a_4$ are much larger for $S=1/2$ than for
$S=1$ and increase with increasing dimerization.
%\begin{table}[t]
%	\includegraphics*[width=.99\linewidth]{DimTab}
	
%\end{table}

Apart from the 
%long range limit 
large-$r$ expansion, an analytical expression for small distances can
also be obtained from Eqs.~(\ref{MSWTCFo}) and
(\ref{MSWTCFe}). % First
%Turning to the short-range case of the uniform
%chain, we note that Eqs.~(\ref{MSWTCFo}) and (\ref{MSWTCFe})
%reduce to
%\begin{equation}\label{SSFuni}
%\left<\vec S_{j+r}\cdot\vec S_j\right>\approx
%\left(\frac{1}{N}\sum_kn_\textnormal{B}(\zeta_k)\cos(kr)\right)^2.
%\end{equation}
%At low temperature the main contribution to the sum stems from the
%points around $k=0$. Thus, for $kr\ll1$ the cosine may be expanded
%to second order. Furthermore using the constraint Eq.~(\ref{MSWTconstraint}) at
%$\delta=0$ as well as expanding the dispersion in the remaining term to
%quadratic order we obtain
%\begin{equation}\label{SRCF}
%\left<\vec S_{j}\cdot\vec
%S_{j+r}\right>\approx\left(S-\frac{t_0^{3/2}
%\textnormal{Li}_{\frac{3}{2}}(\textnormal{e}^{-v_0})}{8\sqrt{\pi}}r^2\right)^2
%\end{equation}
%Inserting the
%reduced chemical potential into Eq.~(\ref{SRCF}) and expanding in
%the regime of low temperature, we obtain
%\begin{equation}
%\left<\vec S_j\cdot\vec S_{j+r}\right>\approx S
%\left(S-\frac{t_0^{3/2}(\zeta(\frac{3}{2})-2\sqrt{\pi v_0}
%-\zeta(\frac{1}{2})v_0)}{4\sqrt{\pi}}r^2\right).
%\end{equation}
%First we witness
%that for sufficiently low temperatures the main contributions to
%the sums occurring in Eqs. (\ref{MSWTCFo}) and (\ref{MSWTCFe})
%stem from the lower magnon branch (compare Fig.~\ref{Oc}). 
Again neglecting contributions stemming from $n_\text{B}(\zeta_k^+)$
and expanding the dispersion as well as the function
$f_k(j,r,\delta)$, Eq.~(\ref{fjr}), to second order in $k$ we find the
short-distance expansion
%\begin{widetext}
\begin{equation}\label{SSsr}
\begin{aligned}
\left<\vec S_j\cdot\vec
S_{j+r}\right>&%\approx\left\{S-\frac{t_{\delta}
%\textnormal{Li}_\frac{3}{2}\left(\textnormal{e}^{-v_{\delta}}\right)}
%{8\sqrt{\pi}}\left[r^2+\delta\left((-1)^r-1\right)
%\left((-1)^jr-\frac{\delta}{2}\right)\right]\right\}^2\\
\approx S\left\{S-\frac{t_\delta^{3/2}(\zeta(\frac{3}{2})
-2\sqrt{\pi v_\delta}-\zeta(\frac{1}{2})v_\delta)}{4\sqrt{\pi}}\right.\\
&\left.\times\left[r^2+\delta\left((-1)^r-1\right)\left((-1)^jr
-\frac{\delta}{2}\right)\right]\right\}.
\end{aligned}
\end{equation}
%\end{widetext}
In Fig.~\ref{LogCF} the SCF is presented for $S=1$, $T/J=0.1$ and
$\delta=0.3$ as a function of distance. Fig.~\ref{LogCF}(a) compares a
numerical self-consistent solution of Eqs.~(\ref{MSWTCFo}) and
(\ref{MSWTCFe}) with the
%short range limit, 
expansion for small distances $r$, Eq.~(\ref{SSsr}), and the long
distance asymptotics, Eq.~(\ref{SSdlr}). Furthermore, numerical TMRG
data are shown. For $r\gg1$ the exponential decay of the SCF, as
derived from the MSWT is clearly visible.  When applicable, we find
good agreement between the analytical expressions for short and long
distances and the results obtained by TMRG. The same quantities for
the same parameters are shown in Fig.~\ref{LogCF}(b) on a linear
scale.  In Fig.~\ref{lndell} we compare the ACF for spin $S=1$ and two
large values of the dimerization parameter, $\delta=0.6$ and $0.9$, as
obtained from MSWT, with TMRG data. Although a high dimerization leads
to a high effective temperature $t_\delta$ the agreement of the MSWT
with the numerical data is still remarkably good.

\begin{figure}
	\includegraphics*[width=.95\linewidth]{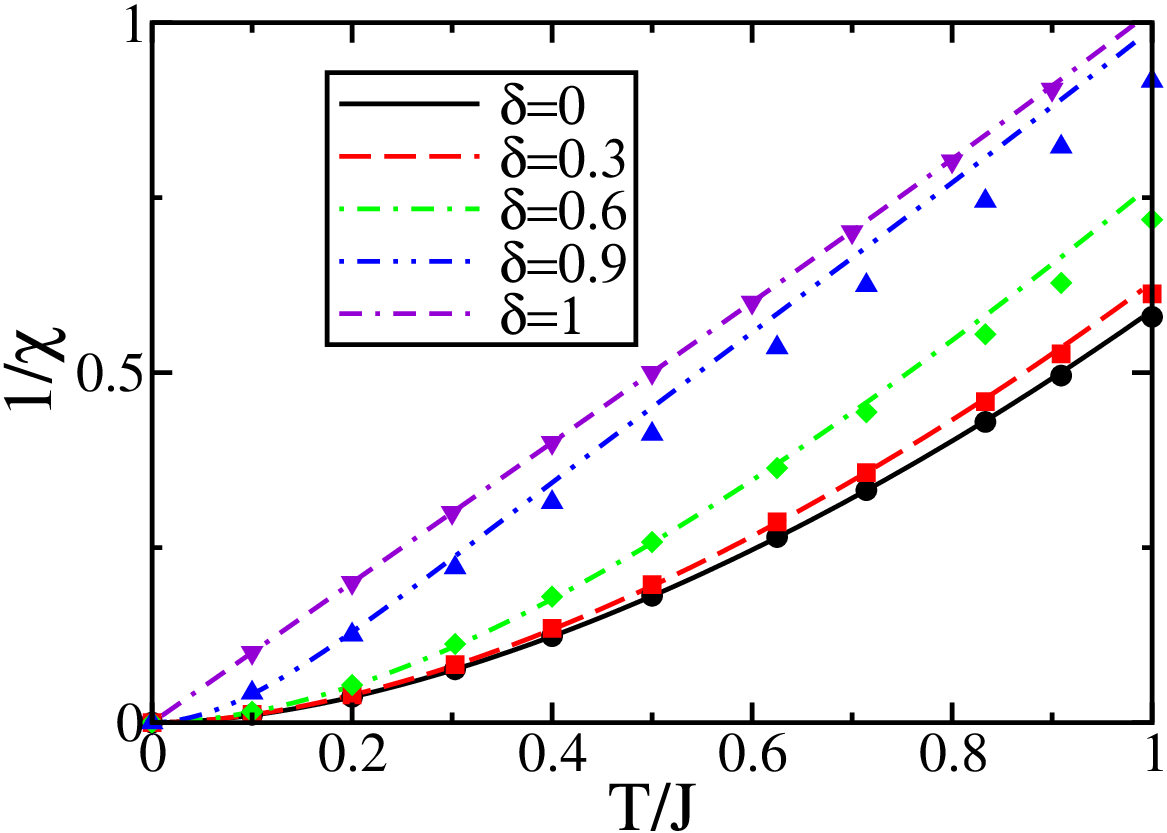}
      \caption{(Color online) Inverse susceptibility $1/\chi$ as a
        function of $T/J$ for $S=1$ and different values of $\delta$
        as indicated.  The lines show the results from MSWT, the
        symbols for $\delta<1$ TMRG data, and the symbols for
        $\delta=1$ the exact solution for decoupled dimers. An
        increasing dimerization suppresses $\chi$, and the decoupled
        dimers follow a Curie law $\chi\sim1/T$.\\\\}\label{1ochi}
	\end{figure}
%%JS

      Next, we calculate the zero field susceptibility per
      site
\begin{equation}\label{chi}
\chi=\frac{1}{3NT}\sum_{j,l}\left<\vec S_j\cdot\vec S_l\right>.
\end{equation}
Applying the Dyson-Maleev transformation we find
\begin{equation}\label{chiMSWT}
\chi=\frac{S}{3T}+\frac{1}{6NT}\sum_k\left\{
(n_k^+)^2+(n_k^-)^2\right\},
\end{equation}
with
$$
	n_k^\pm\equiv n_\text{B}(\zeta_k^-)\pm n_\text{B}(\zeta_k^+).
$$
In order to obtain an analytical result for the susceptibility, we may
interpolate the SCF in Eq.~(\ref{chi}) between the short-range limit
given in Eq.~(\ref{SSsr}) and the asymptotic expression of
Eq.~(\ref{SSdlr}). From this we obtain
\begin{equation}\label{chian}
\begin{aligned}
\chi=\frac{2}{3}\frac{\xi_\delta S^2}{ T}&\left[
1-\frac{3\zeta(\frac{1}{2})}{2\sqrt{\pi}S}
\frac{1}{\sqrt{\xi_\delta}}-\frac{2\pi S
\delta-3\zeta^2(\frac{1}{2})} {4\pi
S^2}\frac{1}{\xi_\delta}+\dots\right],
\end{aligned}
\end{equation}
where we have used that to lowest order the correlation length is
given by $\xi_\delta=S/t_\delta$. In the limit of vanishing
dimerization our result reduces to the result previously found by
Takahashi for the uniform chain.\cite{MSWTlong,chiremark}

In the opposite limit of full dimerization the susceptibility can
easily be calculated exactly.  In this case we have decoupled
ferromagnetic dimers and the suceptibility follows a Curie law at low
temperatures.  Within MSWT we find from Eq.~(\ref{chiMSWT})
\begin{equation}\label{chid1MSWT}
	\chi(\delta=1)=\frac{S+2\{n^2_\text{B}(v_1)+n^2_\text{B}(4JS+v_1)\}}{3T}.
\end{equation}
For $T/J\ll1$ this expression reduces to
\begin{equation}\label{Curie}
	\chi\approx\frac{S(2S+1)}{3T},
\end{equation}
which is the Curie law (per spin) for decoupled ferromagnetic dimers.
In Fig. \ref{1ochi} the inverse susceptibility $1/\chi$ as obtained
from MSWT is compared with TMRG data. The agreement is good for
$t_\delta/S\leq1$ and also the case of decoupled dimers is accurately
captured by MSWT.

\section{Dynamical spin structure factor}
\label{sec:dyn}

For the dimerized AF Heisenberg chain the dynamical spin structure factor (dynamic SSF)
has been intensely studied by various methods.\cite{UhrigSchulz,PoilblancRiera,AugierPoilblanc,WeiqiangHaas}
% For sufficiently small dimerization the well known Des Cloiseaux - Pearson excitation spectrum\cite{CloiseauxPearson} 
% of the AFM Heisenberg chain is obtained. However even small dimerizations can give rise to distinct features. 
% For instance for $\delta=0.048$, relevant for the spin Peierls compound NaV$_2$O$_5$, the intensity in the DSSF 
% is zero if the energy does not exceed the spin gap.\cite{WeiqiangHaas}
Here we will discuss the impact of a finite dimerization on the dynamic SSF
of the FM Heisenberg chain.  The dynamical properties of the uniform
1D ferromagnet within MSWT have been adressed by
Takahashi.\cite{MSWT2,MSWT3} In this case for a given value of $q$ the
dynamic SSF exhibits a magnon peak from which the magnon dispersion can be
obtained. In this approximation magnon-magnon interactions are not
taken into account and the peaks are broadened due to thermal
fluctuations only. The magnon peaks, which exhibit a Lorenzian
lineshape at finite temperatures, thus reduce to $\delta$-functions at
$T=0$. Moreover, there is a two-magnon continuum beyond which the dynamic SSF
vanishes.  New aspects to this picture have been added
recently\cite{Her11} when it was shown that an edge singularity occurs
at the boundary of the two-magnon continuum caused by a diverging
density of states.

Here we want to generalize this analysis to the dimerized FM chain.
Within the standard approach of neutron scattering the differential cross-section
is directly related to the dynamic SSF.\cite{Lovesey} The latter can be obtained by a Fourier
transform of the two-point correlation function. It was shown in the context
of X-ray scattering that further insight into the microscopic properties
of physical systems may be achieved if the incident photon fulfills the Bragg
condition such that the resulting state is given by
a coherent superposition of the forward diffracted and the Bragg reflected
wave.\cite{Schuelke1,Schuelke2,Schuelke3} The structure factor
can be written as a matrix in terms of the reciprocal lattice vectors
in analogy to the dielectric matrix whose inverse
defines the dynamic charge structure factor of a dielectric.\cite{Adl62,Wis63,Lin88}

%	S_{jl}(q,
%	G,\omega)=\delta_{jl}S(q,q,\omega)+(1-\delta_{jl})S(q,q+G,\omega)
%
%where the diagonal elements give the DSSF obtained without Bragg reflection and the off-diagonal elements 
%take account of the difference of transfered momenta for the forward diffracted and the Bragg reflected photon.

In the same spirit we formulate the dynamic SSF of the dimerized FM chain as a $2\times2$-matrix 
% where the
% commensurability with $k=\pi$ of the modulation of the interaction strength (see Hamiltonian
% (\ref{H})) leads to
$$
	\vec S(q,\omega)=\left(
	\begin{array}{cc}
		S(q,q,\omega)&S(q,q+\pi,\omega)\\
		S(q+\pi,q,\omega)&S(q+\pi,q+\pi,\omega)
	\end{array}
	\right).
$$
In order to obtain this matrix we calculate the Green's function
\begin{equation}\label{G}
	G(j,r,\tau)\equiv-\left<\mathcal{T}\left[\vec{S}_j(0)\cdot\vec{S}_{j+r}(\tau)\right]\right>,
\end{equation}
which within MSWT reads %For this reason, we consider the following
%expressionß
%
%\begin{equation}\label{Gtau}
%G_{\delta}(j,r,\tau)=
%\begin{cases}
%%G_{\delta}^\textnormal{ee}(j,r,\tau)\quad j\textnormal{ even, }r
%\textnormal{ even},\\
%G_{\delta}^\textnormal{eo}(j,r,\tau)\quad j\textnormal{ even, }r
%\textnormal{ odd},\\
%G_{\delta}^\textnormal{oe}(j,r,\tau)\quad j\textnormal{ odd, }
%r\textnormal{ even},\\
%G_{\delta}^\textnormal{oo}(j,r,\tau)\quad j\textnormal{ odd, }
%r\textnormal{ odd}.\\
%\end{cases}
%\end{equation}
%
\begin{equation}
\begin{aligned}
G(j,r,\tau)&=
-\left<\mathcal{T}\left\{a^{\ }_{j}(0)a_{j+r}^\dagger(\tau)\right\}\right>\\
&\times\left<\mathcal{T}\left\{a_{j}^\dagger(0) a^{\ }_{j+r}(\tau)\right\}\right>.
\end{aligned}
\end{equation}
From this the matrix elements of the dynamic SSF can be obtained straightforwardly via
\begin{equation}
	S(q,q',\omega)=2n_\text{B}(-\omega)\text{Im}G^\text{ret}(q,q',\omega),
\end{equation}
where $G^\text{ret}(q,q',\omega)$ is the retarded finite temperature Green's
function obtained from $G(j,r,\tau)$. For the diagonal elements we find $S(q,q,\omega)=S(q+\pi,q+\pi,\omega)$ with
\begin{equation}\label{DSSF}
	\begin{aligned}
		S(q,q,\omega)=&\frac{\pi}{N}n_\text{B}(-\omega)\sum_{\sigma,\sigma'\in\{\pm\}}\sum_k
		\delta(\omega-\epsilon^{\sigma\sigma'}_q(k))\\
		&\times\left\{1+\sigma\sigma'\Phi(k,q+k)\right\}\mathcal{N}[\zeta_k^\sigma,
		\zeta_{q+k}^{\sigma'},\beta],
	\end{aligned}
\end{equation}
where we have defined
\begin{equation}\label{Ratio}
	\Phi(k,k')\equiv\frac{\cos k\cos k'-\delta^2\sin k\sin k'}
	{\sqrt{1-(1-\delta^2)\sin^2k}\sqrt{1-(1-\delta^2)\sin^2k'}}.
\end{equation}
and%
\begin{equation}
\begin{aligned}
\mathcal{N}[\zeta^\sigma_k,\zeta^{\sigma'}_{q+k},\beta]\equiv&
n_\textnormal{B}(\zeta_{k}^{\sigma})\{1+n_\textnormal{B}(\zeta_{q+k}^{\sigma'})\}\\
&\times\left(\textnormal{e}^{-\beta\epsilon_q^{\sigma\sigma'}(k)}-1\right)\,,
\end{aligned}
\end{equation}
with the two particle continuum 
$\epsilon_q^{\sigma\sigma'}(k)\equiv\zeta_k^\sigma-\zeta_{q+k}^{\sigma'}$.
As in the case of the uniform 1D ferromagnet the dynamic SSF for the dimerized chain
fulfills detailed balance.
\begin{figure}[!t]
\includegraphics[angle=-0,scale=.305]{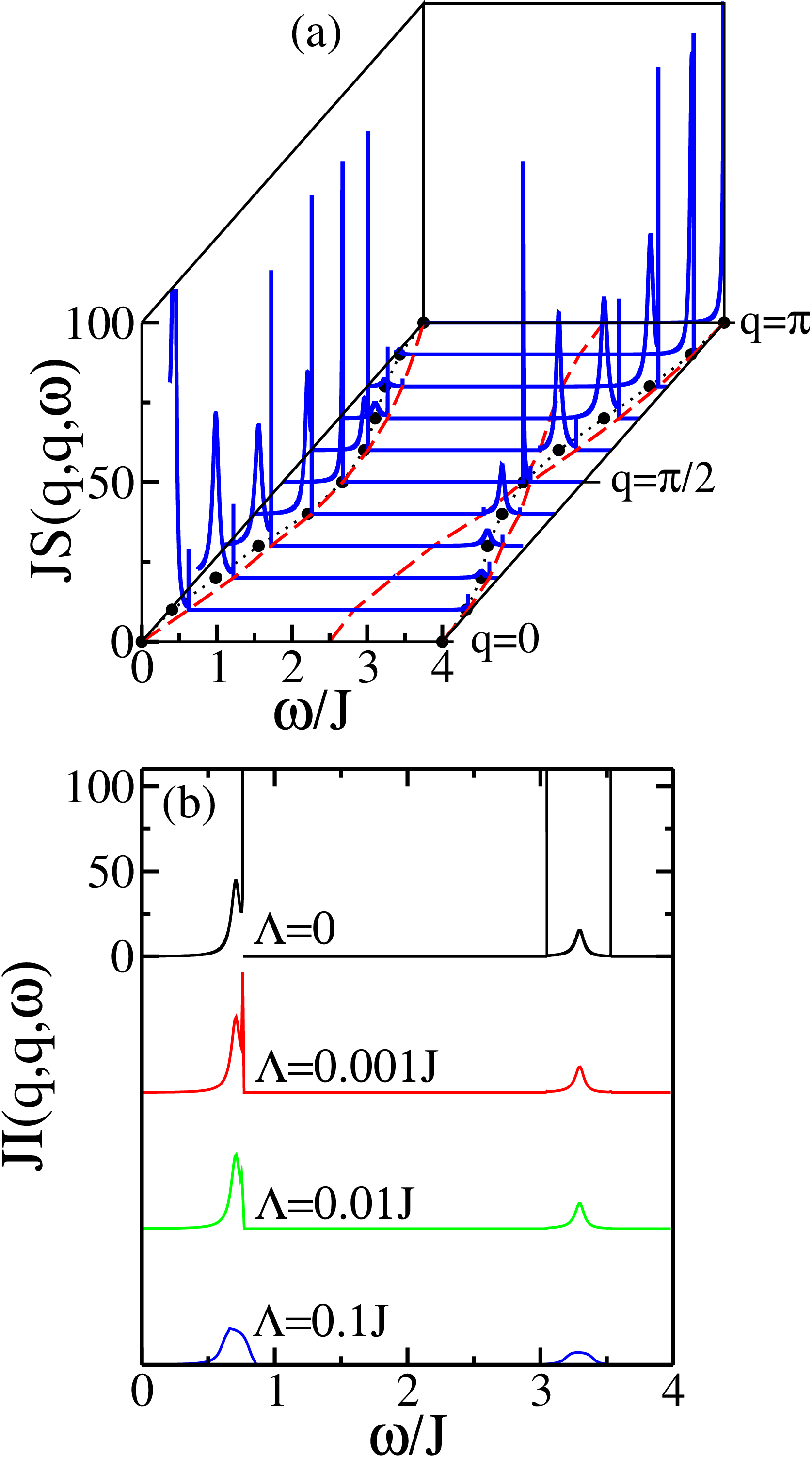}
\caption{(Color online) Diagonal elements of the dynamic spin structure factor 
$S(q,q,\omega)$ for a dimerized ferromagnetic chain and $S=1$, $T/J=0.1$: 
(a) $S(q,q,\omega)$ spectra for different momenta $q$ and
$\delta=0.6$. The red dashed lines show the edge boundaries and
the symbols show the peak positions projected onto the
$(q,\omega)$-plane. (b) broadened spectra $I_{\Lambda}(q,q,\omega)$ for $q=4\pi/5$,
%%$I_\Lambda(q=4\pi/5,q=4\pi/5,\omega)$ , 
$\delta=0.4$ and for different $\Lambda=0, 0.001, 0.01,$ and $0.1 J$.
Vertical lines in  $\Lambda=0$ panel  indicate positions of edge singularities.}\label{SFTp1dp6}
\end{figure}

\begin{figure}[t]
\includegraphics[angle=-0,scale=.305]{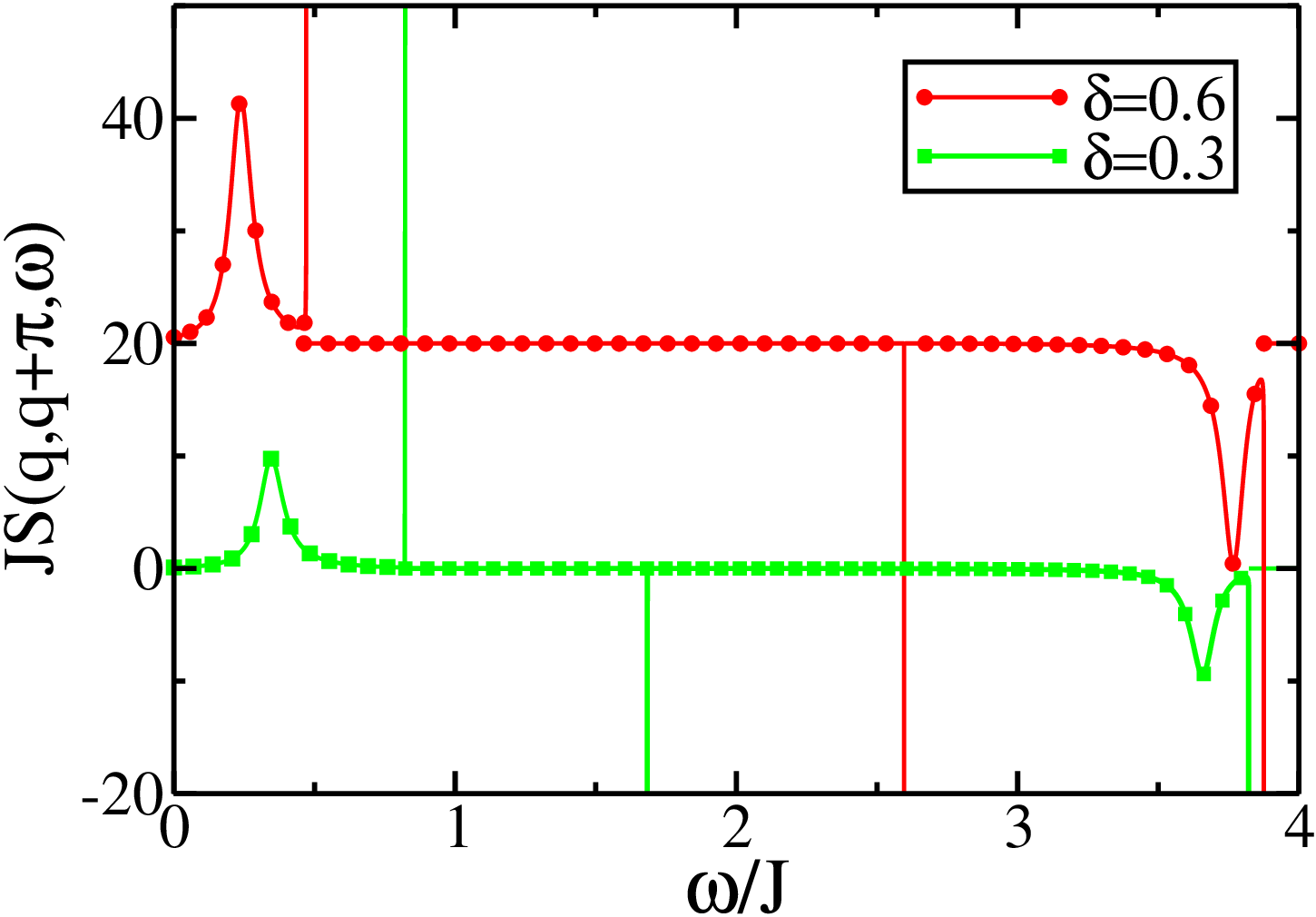}
\caption{(Color online) Off-diagonal element of the dynamical spin
  structure factor matrix $S(q,q+\pi,\omega)$ for $q=\pi/5$, $S=1$,
  $T/J=0.1$, and $\delta$ as indicated. Note that the data for
  $\delta=0.6$ are shifted by 20 units along the ordinate for clarity
  of presentation. The vertical lines indicate positions of edge
  singularities (see text).}
\label{DSP}
\end{figure}

In this approximation the two magnon continuum determines the regions
where $S(q,q,\omega)$ is nonzero. % and there may
% exist edge singularities. 
This can be seen in Fig.~\ref{SFTp1dp6}(a) where the diagonal element
of the dynamic SSF for the dimerized chain is shown for $\delta=0.6$. The
symbols reflect the peak positions projected onto the
$(q,\omega)$-plane. They follow the reduced magnon dispersions
$\zeta_q^\pm$. Since we are dealing with two separate magnon branches,
an edge singularity occurs at each boundary of the two magnon continua due to the divergence of the respective density of states. The
boundaries beyond which the dynamic SSF for the respective branch vanishes
are given by the red dashed lines. For $\omega>0$ we find three
boundaries out of the four possible combinations because for
$\epsilon_q^{-+}$ the argument of the delta function in
Eq.~(\ref{DSSF}) never vanishes.

%%JS
Experimentally, the energy resolution in a measurement is always
finite. An important question to ask is therefore if the edge
singularities can be resolved at all. We address this question by
studying the dynamical SSF with a finite energy resolution $\Lambda$
$$
	I_\Lambda(q,q,\omega)=\frac{1}{2\Lambda}\int\limits_{\omega-\Lambda}^{\omega+\Lambda} d\omega_0\ S(q,q,\omega_0).
$$
Results for this quantity are shown in Fig.~\ref{SFTp1dp6}(b) for
$\delta=0.4$ and various resolutions. Being very close to a magnon
peak, the lower edge singularity has a substantial spectral weight.
Even with a relatively moderate energy resolution this edge
singularity is therefore detectable. On the contrary, the two edge
singularities at higher energies carry almost no spectral weight in
the considered example making them experimentally irrelevant.

The off-diagonal elements of the dynamic SSF,
$S(q,q+\pi,\omega)=S(q+\pi,q,\omega)$, are given by
\begin{equation}\label{DSPE}
	\begin{aligned}
		S(q,q+\pi,\omega)=&\frac{\pi}{N}n_\text{B}(-\omega)\sum_{\sigma,\sigma'\in\{\pm\}}\sum_k\delta(\omega-\epsilon^{\sigma\sigma'}_q(k))\\
		&\times\sigma\sigma'\Psi(k,q+k)\mathcal{N}[\zeta_k^\sigma,\zeta_{q+k}^{\sigma'},\beta],
	\end{aligned}
\end{equation}
where we have defined
\begin{equation}\label{Psi}
	\Psi(k,k')=\frac{2\delta\cos k\sin k'}{\sqrt{1-(1-\delta^2)\sin^2k}\sqrt{1-(1-\delta^2)\sin^2k'}}.
\end{equation}
Examples for the off-diagonal element of the dynamic SSF matrix are
shown in Fig.~\ref{DSP}. As in the case of $S(q,q,\omega)$ the
two-magnon continuum determines the region where
$S(q,q+\pi,\omega)\neq0$ with edge singularities at the boundaries.
Moreover we find magnon peaks of different signs following the magnon
dispersions.

%Furthermore we note that contrary to the uniform case the static structure factor of the dimerized FM chain does not necessarily decrease monotonically for increasing $|q|$. Instead, for $\delta_S>0$ we observe a distinct structure. This is shown in Fig. \ref{StatFac} where the static structure factor is plotteed for different values of $\delta_S$. First, common to all static structure factors the central peak at $q=\pi$ is broadend. Furthermore, at intermediate dimerization we witness that further maxima arise at higher momenta. However at strong dimerization the static structure factor is very broadened, such that these peaks are lost.

\section{Static spin-structure factor}
\label{sec:sta}
\begin{figure}[t]
	%\begin{minipage}{.48\linewidth}
	\includegraphics[width=.98\linewidth]{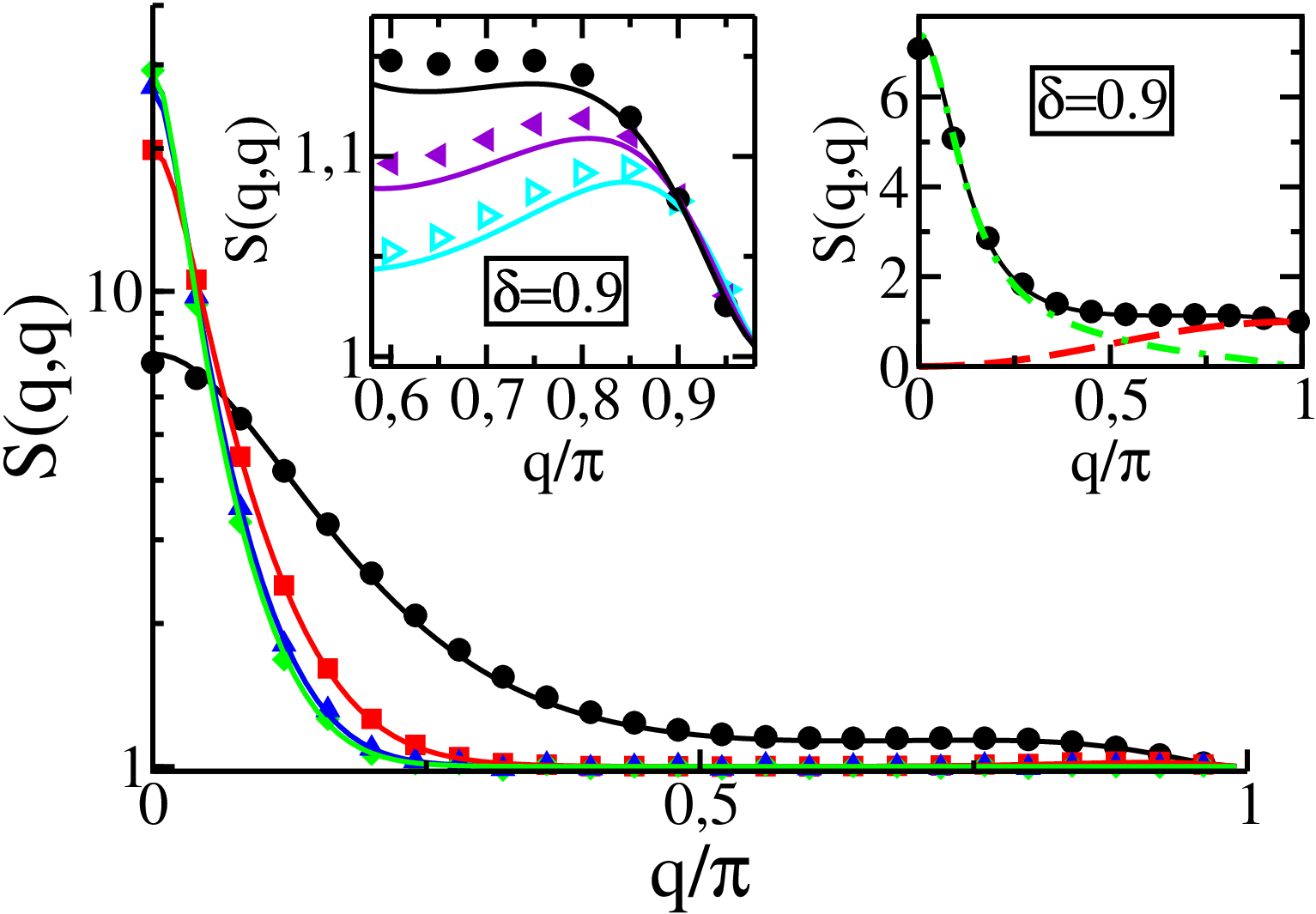}
      \caption{(Color online) Diagonal static spin structure factor
        for $S=1$, $T/J=0.1$. Symbols denote data from TMRG, lines
        results from MSWT. Main: Data for dimerizations $\delta=0$
        (diamonds), $\delta=0.3$ (triangles), $\delta=0.6$ (squares),
        and $\delta=0.9$ (circles). Left inset: High-momentum peak
        for $\delta=0.9$ at $T/J=0.1$ (circles), $T/J=0.08$ (filled
        triangles), and $T/J=0.0625$ (open triangles). Right inset:
        Data for $\delta=0.9$ with contributions stemming from the
        lower and upper branch shown by dashed-dotted and dashed
        lines, respectively.}\label{Sq}
	%	\end{minipage}
	%	\hspace*{.01\linewidth}
	%	\begin{minipage}{.48\linewidth}
	%		\includegraphics[width=.98\linewidth]{StaticStrcutrueFactoranalytic.eps}
	%		\caption{(Color online) Static spin structure factor for $S=1$ and $T/|J|$. We show a 
	%		comparison of numerical results of the MSWT, as given by 
	%		Eq.~(\ref{SSFD}) for $\delta=0$ (circles), $\delta=0.3$ (squares), $\delta=0.6$
	%		(diamonds), and $\delta=0.9$ (triangles) with the respective analytical
	%		expressions (lines). For small momenta the analytical expression
	%		Eq.~(\ref{Sqsq}) gives good agreement with the numerical results. Inset:
	%		High momentum peak position $q_\text{max}$ obtained from a numerical
	%		solution of Eq.~(\ref{SSFD}) for $S=1$ and $\delta=0.6$ (squares),
	%		$\delta=0.7$ (diamonds), $\delta=0.8$ (circles), and $\delta=0.9$
	%		(triangles) as a function of temperature. The straight lines are the
	%		high momentum peak positions as predicted from the analytical result
	%		Eq.~(\ref{qmaxl}).}\label{SSSFana}
	%	\end{minipage}
	\end{figure}
%\begin{figure}[t!]
%\includegraphics[scale=.3]{StatFac.eps}
%\caption{Static structure factor for various values of
%dimerizations. (a) Static spin-structure factor for the uniform
%case (black straight line), $\delta_S=0.3$ (dashed red line),
%$\delta=0.6$ (dashed-dashed-dotted line), and $\delta=0.9$ (blue
%dashed-dotted-dotted line) for $S=1$ and $T/|J|=0.1$. (b)
%Comparison of the static spin-structure factor as obtained from
%the Eqs. (\ref{SSFUNI}) and (\ref{SSFuniapp}) in the uniform case
%as well for $\delta=0.3$ ($\delta=0.6$) as given by the Eqs.
%(\ref{SSFD}) and (\ref{SSFdimapp}) for the same spin quantum
%number and temperature as in (a).} \label{StatFac}
%\end{figure}

The static spin structure factor (static SSF) matrix can be easily obtained from the dynamic SSF via
\begin{equation}\label{SSSFInt}
	 S(q,q')=\int\limits_{-\infty}^\infty\frac{d\omega}{2\pi} S(q,q',\omega).
\end{equation}
We first want to discuss the diagonal elements of this matrix. We divide
% \begin{equation}\label{SSSF1}
$ S(q,q)=S^-(q,q)+S^+(q,q) $
% \end{equation}
with $S^-(q,q)$ ($S^+(q,q)$) being the part which stems from the lower
(upper) branch of the dispersion. These contributions are given by
Eqs.~(\ref{DSSF}) with $\sigma=\sigma'$ and $\sigma\ne\sigma'$ (see
also Fig.~\ref{SFTp1dp6}). Performing the frequency integrals we find
\begin{equation}\label{SSSFl}
\begin{aligned}	S^-(q,q)=\frac{1}{2N}&\sum_{\sigma\in\{\pm\}}\sum_k\left[1+\Phi(k,q+k)\right]n_\text{B}(\zeta^\sigma_k)\\&\times\left[1+n_\text{B}(\zeta_{q+k}^\sigma)\right],
\end{aligned}
\end{equation}
and
\begin{equation}\label{SSSFu}
\begin{aligned}	S^+(q,q)=\frac{1}{2N}&\sum_{\sigma\in\{\pm\}}\sum_k\left[1+\Phi(k,q+k)\right]n_\text{B}(\zeta^\sigma_k)\\&\times\left[1+n_\text{B}(\zeta_{q+k}^{-\sigma})\right].
\end{aligned}
\end{equation}
From this we have
\begin{equation}\label{SSFD}
S(q,q)=\frac{1}{2N}
\sum_k\Biggl\{n_k^+n_{q+k}^++n_k^-n_{q+k}^-\Phi(k,q+k)
\Biggr\}+S.
\end{equation}
Alternatively, the diagonal elements of the static SSF can be obtained
directly from
\begin{equation}\label{SqSCF}
	S(q,q)=\frac{1}{N}\sum_{j,l}\left<\vec S_j\cdot\vec S_l\right>\text{e}^{iq(j-l)}.
\end{equation}
However, within the latter approach the distinction of the
contributions stemming from the lower and the upper branch of the dynamic SSF
is less transparent.

In Fig.~\ref{Sq} the diagonal structure factor $S(q,q)$ is
shown % as a function of q
for various dimerizations at $T/J=0.1$ with $S=1$. The agreement
between the MSWT results and the numerical TMRG data is good even for
high dimerizations. An increasing dimerization leads to an increase of
the linewidth of the central peak at $q=0$. At high dimerizations a
shoulder in $S(q,q)$ appears that extends almost up to $q=\pi$. At low
temperatures a second maximum develops near $q=\pi$
%%JS
(see left inset of Fig.~\ref{Sq}). For decreasing temperatures the
peak shifts to higher momenta leading to a sharp maximum close to the
boundary of the Brillouin zone for very low temperatures and high
dimerizations.

For a dimerized FM chain with $\delta<1$ such a high
momentum structure is expected due to the commensurate modulation of the exchange
constant.\cite{remarkd1} 
The naive expectation of a peak right at $\pi$ indicating the dimerization
of correlation functions is, however, not fulfilled.
%However from this picture one would expect a high
%momentum peak at $q=\pi$. 
% Interestingly we find a shifted high momentum peak.
Instead, we find a small peak with momentum shifted away from $q=\pi$.
This can be readily understood by considering $S^-(q,q)$ and
$S^+(q,q)$ separately as is done in the right inset of Fig.~\ref{Sq}.
$S^-(q,q)$ is peaked at $q=0$ and $S^+(q,q)$ reveals a maximum at
$q=\pi$. However, at sufficiently high dimerizations the $S^-(q,q)$
contribution also yields a significant contribution for $q\ge\pi/2$.
Therefore the two contributions add up to a peak which is shifted with
respect to $q=\pi$.
% We emphasize that for small dimerization the contribution
% from the lower branch of the diagonal elements of the DSSF matrix is
% much smaller and hence the $S(q,q)$ decreases monotonously (Note that
% $S(q,q)\geq1$).

One may also wonder why the dimerization does not show up more
pronounced in the diagonal elements of the static spin structure
factor matrix, i.e., why the peak at $q=0$ has much more weight than
the high momentum peak even at strong dimerizations. To investigate
this property, we insert Eq.~(\ref{SSdlr}) into Eq.~(\ref{SqSCF})
dividing the SCF into a uniform and an alternating part. Performing
the sum over $r$ to lowest order we end up with
\begin{equation}
	S(q,q)\approx\frac{1}{\xi_\delta}\left\{\frac{A}{q^2+\xi_\delta^{-2}}+\frac{B}{(q-\pi)^2+\xi^{-2}_\delta}\right\},
\end{equation}
with
$$
	A=\frac{t_\delta}{4}\left(\frac{1}{v_\delta}+\delta^2t_\delta\right)
$$
and
$$
	B=\frac{t_\delta^2\delta^2}{4}.
$$
Thus, as expected, we find two Lorentzians at $q=0$ and $q=\pi$,
respectively. However, the ratio
\begin{equation}
\label{ratio}
	\frac{S(\pi,\pi)}{S(0,0)}=\frac{v_\delta t_\delta\delta^2}{1+v_\delta t_\delta\delta^2}
\end{equation}
is small for effective temperatures $t_\delta\ll 1$. For large
$t_\delta$, on the other hand, we have $\xi_\delta\ll 1$ and the two
Lorentzians become very broad.
%\cite{remarkSq}

Finally, from the left inset of Fig.~\ref{Sq} we see that the peak
position of the maximum at high momentum is temperature dependent. On
the other hand, we find that for $t_\delta$ and $S$ fixed the peak
position does not depend on $\delta$. To understand this puzzling
feature we derive an analytical expression for the position of the
high momentum peak. Keeping only contributions from
$n_\text{B}(\zeta_k^-)$ in Eq.~(\ref{SSFD}), expanding $\Phi(q,k)$
around $q=\pi$ and $k=0$, and performing a saddle-point integration we
can evaluate the remaining expression. Using Eq.~(\ref{van}) we find that the
high-momentum peak is located at
\begin{equation}\label{qmaxl}
	q_\text{max}=\pi-\sqrt[4]{\frac{2}{S^2}}t_\delta^{3/4}+\frac{1}{2}\frac{t_\delta^{5/4}}
	{\sqrt[4]{8S^6}}+\dots
\end{equation}
which explains why the peak position does not depend explicitly on
$\delta$. % for $t_\delta$, $S$ fixed.

\begin{figure}
	%\begin{minipage}{.48\linewidth}
	\includegraphics[width=.98\linewidth]{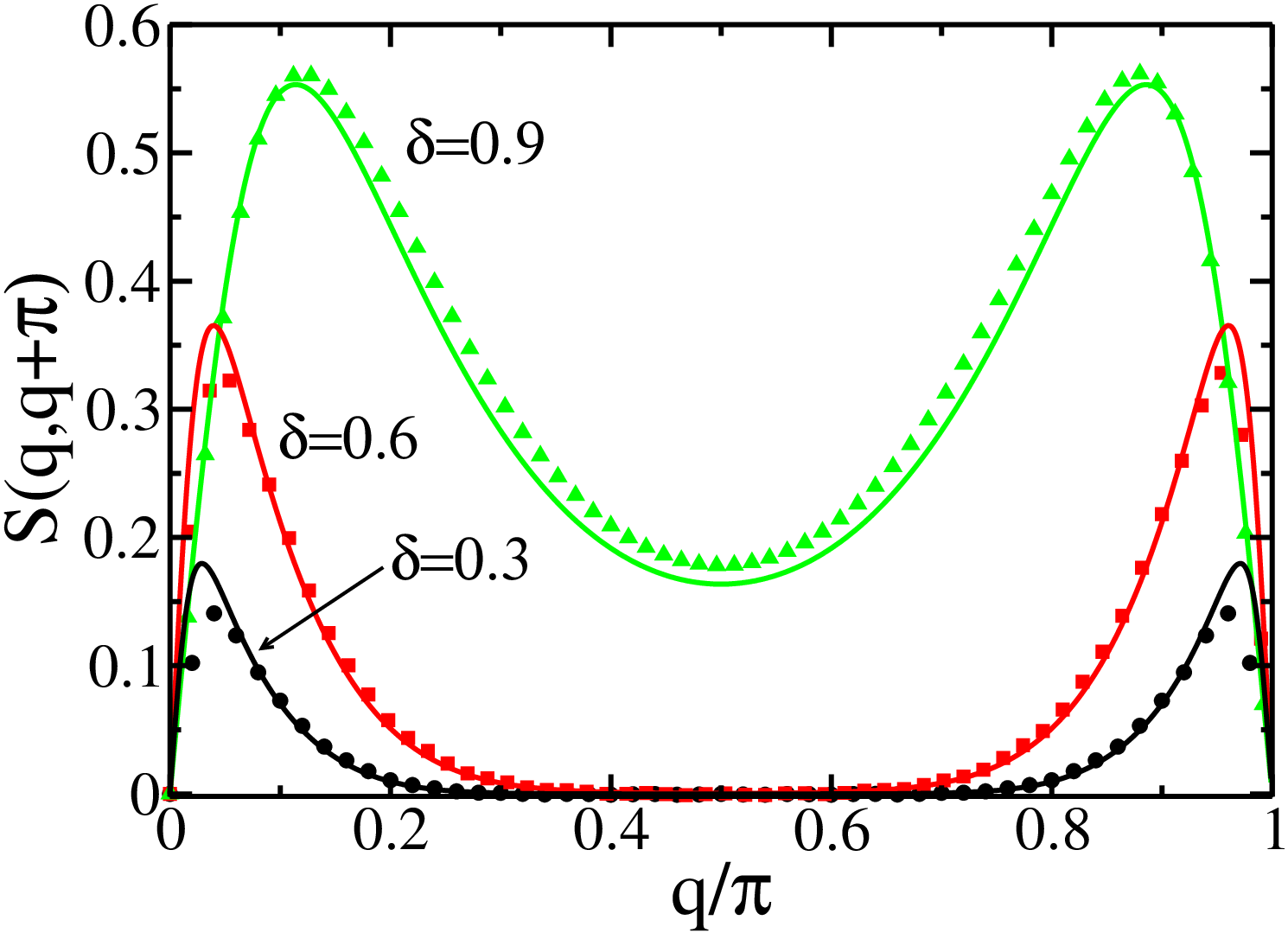}
		\caption{(Color online) Off-diagonal elements of the static spin structure factor matrix for $S=1$, $T/J=0.1$,
		and various dimerizations as a function of $q$. The symbols 
% (triangles: $\delta=0.9$, squares: $\delta=0.6$, and circles:	$\delta=0.3$) 
              show numerical data obtained by TMRG, whereas the
              solid lines are results from MSWT, Eq.~(\ref{SSPE}).}\label{PSSSF}
	%	\end{minipage}
	%	\hspace*{.01\linewidth}
	%	\begin{minipage}{.48\linewidth}
	%		\includegraphics[width=.98\linewidth]{StaticStrcutrueFactoranalytic.eps}
	%		\caption{(Color online) Static spin structure factor for $S=1$ and $T/|J|$. We show a 
	%		comparison of numerical results of the MSWT, as given by 
	%		Eq.~(\ref{SSFD}) for $\delta=0$ (circles), $\delta=0.3$ (squares), $\delta=0.6$
	%		(diamonds), and $\delta=0.9$ (triangles) with the respective analytical
	%		expressions (lines). For small momenta the analytical expression
	%		Eq.~(\ref{Sqsq}) gives good agreement with the numerical results. Inset:
	%		High momentum peak position $q_\text{max}$ obtained from a numerical
	%		solution of Eq.~(\ref{SSFD}) for $S=1$ and $\delta=0.6$ (squares),
	%		$\delta=0.7$ (diamonds), $\delta=0.8$ (circles), and $\delta=0.9$
	%		(triangles) as a function of temperature. The straight lines are the
	%		high momentum peak positions as predicted from the analytical result
	%		Eq.~(\ref{qmaxl}).}\label{SSSFana}
	%	\end{minipage}
	\end{figure}
Next, we turn to the off-diagonal elements of the static SSF matrix. 
From integrating Eq.~(\ref{DSPE}) over frequency we find
\begin{equation}\label{SSPE}
	S(q,q+\pi)=\frac{1}{2N}\sum_k\Psi(k,q+k)n_k^-n_{q+k}^-.
\end{equation}
The off-diagonal elements vanish in the undimerized case and are finite
in the dimerized case.
Thus they
provide a rigorous criterion for dimerized spin correlation functions.  
In Fig.~\ref{PSSSF} this expression is shown for $S=1$, $T/J=0.1$, and
various dimerizations.
%%\cite{remarkPSSSF}
%%PH
$S(q,q+\pi)$ is an odd function of momentum transfer $q$, and is a symmetric function
with respect to $q=\pi/2$. It exhibits a two peak structure which is
more pronounced the higher the dimerization is. In complete analogy to
the high-momentum peak of $S(q,q)$ we can again determine the
positions of the maxima of $S(q,q+\pi)$ and find
\begin{equation}\label{qmax2}
	q_\text{max}=\frac{\pi}{2}\pm\left(\frac{1}{\xi_\delta}-\frac{\pi}{2}\right).
\end{equation}
Hence, as in the previous case, the peak positions of $S(q,q+\pi)$ are
determined by the spin quantum number $S$ and the reduced temperature
$t_\delta$ only. In particluar we observe that the distance from the
center (boundary) of the Brillouin zone of the low (high) momentum
peak is given by the inverse correlation length $\xi_\delta^{-1}$.

To summarize, dimerization is signaled by the appearance of finite
intensity in the off-diagonal spin structure factor. On the other
hand, the conventional (diagonal) structure factor reflects the FM
spin-Peierls dimerization mainly by quantitative changes, in
particular by an increase of the linewidth of the $q=0$ peak. As a
qualitative change, a high-momentum maximum at large dimerizations and
low-temperatures develops in the diagonal structure factor, however,
this structure always has a very small intensity.

\section{Conclusions}
\label{sec:sum}
We have shown that an alternation of the magnetic exchange can also
occur in ferromagnetic spin chains. Possible mechanisms are a coupling
to orbital fluctuations, as in
YVO$_3$\cite{PhysRevLett.91.257202,SirkerKhaliullin,SPPRL}, or to
lattice degrees of freedom. In both cases the dimerization has to be
thermally activated leading to a gain in magnetic energy $\sim
-T^{3/2}\delta^2$. In an adiabatic approximation the loss in potential
energy is $\sim K\delta^2$ where $K$ is an effective elastic constant.
A dimerized phase due to coupling to the lattice is therefore only
possible as a finite-temperature phase and only if $K$ is smaller than
some threshold value.
% We suggest that monatomic wires on
% surfaces\cite{Gambardella}, where chain atoms are relatively weakly
% bound, might be possible candidates.

As an analytical method to calculate the thermodynamic properties of a
dimerized ferromagnetic Heisenberg chain we have used a modified
spin-wave theory. By comparing with numerical data obtained by the
transfer matrix renormalization group, we have shown ---as a central
result of this paper---that the modified spin-wave theory yields
excellent results for a wide range of dimerization parameters and
temperatures.

We found that both the spin and the dimer correlation functions decay
exponentially at finite temperatures with the same correlation length
$\xi_{\delta} = c(T,\delta)
JS^2/T$. %%$\xi_{\delta}$ as given by Eq.~(\ref{Xi}).
The correlation length $\xi_{\delta}$ decreases with increasing
dimerization $\delta$.
%and the faster the larger the dimerization is. 
Next-leading corrections to the $1/T$ behavior have been calculated
explicitly for spin $S=1/2$ and $S=1$ and even these corrections (of
order 1) described by the coefficient $c(T,\delta)$ have been shown to
be in excellent agreement with the numerical data. For dimerizations
$\delta<1$ the magnetic susceptibility diverges as $\chi\sim
(1-\delta^2)/T^2$, i.e., the susceptibility at low temperatures is
suppressed with increasing dimerization. For $\delta=1$ we have a
system of decoupled dimers and $\chi\sim 1/T$.

Similarly to X-ray scattering experiments, where the incident photon
is given by a coherent superposition of the forward diffracted and the
Bragg reflected wave, we have formulated the dynamic spin stucture
factor for neutron scattering experiments as a matrix. The diagonal
elements correspond to the standard dynamic spin structure
factor %% in the absence of Bragg reflections and
while the off-diagonal elements include a reciprocal lattice vector.
% for the
% forward diffracted and the Bragg reflected wave. 
The dimerization leads to two magnetic excitation branches which are
clearly visible in the diagonal elements of the dynamic spin structure
factor matrix.  Within the spin-wave approximation used here, this
quantity is zero outside the two-magnon continuum. At the boundary of
the two-magnon continuum edge singularities are formed due to a
diverging density of states.  These singularities might become smooth
once higher magnon excitations are included but we expect that a
peak-structure remains along the two-magnon boundary. 
%%JS
Experimentally,
the edge singularities can only be resolved if they carry sufficient
spectral weight. This is only the case if they occur close to a magnon
peak. 
For the off-diagonal elements of the dynamic spin structure
factor matrix we also find magnon peaks with different signs for the
upper and lower magnon branch. The two-magnon continuum determines the
region where the off-diagonal elements of the dynamic spin structure
factor is nonzero with edge singularities occuring at the boundaries.

The diagonal elements of the static spin structure factor matrix
exhibit $q=0$ peaks with Lorentzian lineshape. The linewidth of the
latter increases with increasing dimerization.  Furthermore, for strong
dimerizations a high momentum peak is observed.  % This peak stems
% from the strong alternation of the spin correlation function and can
% be traced back to the distinct contributions from the lower and upper
% branches of the diagonal elements of the dynamical spin structure factor
% matrix.  
We find that the position of the peak at high momenta depends solely
on $S$ and the reduced temperature  $t_\delta=T/(JS(1-\delta^2))$. %% $t_\delta$ %%defined in Eq.~(\ref{td}).
% Due to
% the exponential decay of all correlation functions there are no Bragg
% peaks associated with the broken symmetry. 
The dimerization leads to a doubling of the unit cell which manifests
itself by the appearance of a finite off-diagonal structure factor.
The off-diagonal elements of the static spin structure factor matrix
show a two-peak structure which becomes more pronounced the higher the
dimerization is. The distance of the maxima with respect to the center
and boundary of the Brillouin zone are given by the inverse
correlation length.

To conclude, a spin-Peierls state can be realized in ferromagnetic
spin chains but only at finite temperatures. 
% Both the gain in magnetic
% energy and the loss of potential energy are $\sim\delta^2$ in the
% adiabatic approximation. Which term wins therefore depends on the
% microscopic parameters of the system. 
Most promising candidates for
this new state are, 
%in our opinion,
as we have argued, monatomic chains on surfaces where the effective
elastic constants are expected to be rather small.

%%%%%%%%%%%%%%%%%%%%%%%%%%%%%%%%%%%%%%%%%%%%%%%%%%%%%%%%%%%%
%%                    Acknowledgments
%%%%%%%%%%%%%%%%%%%%%%%%%%%%%%%%%%%%%%%%%%%%%%%%%%%%%%%%%%%%
\acknowledgments
The authors thank Andres Greco for the careful reading of the manuscript
and useful discussions.
A.M.O. acknowledges support by the Foundation for Polish Science (FNP)
and by the Polish Ministry of Science and Higher Education under
Project N202 069639. J.S. acknowledges support by the MAINZ (MATCOR)
school of excellence and the DFG via the SFB/Transregio 49.


\begin{thebibliography}{51}
\bibitem[{ene()}]{energy}
\bibinfo{note}{The dimerization parameter $\delta$ is proportional to the
deviation from the equillibrium distance of two neighboring sites. Note that
the modulation is comensurate with a wave vector $k=\pi$.}
\bibitem[{chi()}]{chiremark}
\bibinfo{note}{Note that in Takahashi's work the suceptibility was defined with
an additional prefactor of 4.}
\bibitem[{rem({\natexlab{a}})}]{remarkd1}
\bibinfo{note}{We note that for the case of decoupled dimers, $\delta=1$,
Eq.~(\ref{SSFD}) is trivially evaluated. For sufficiently low temperatures
MSWT yields $S(q)=S(S+1)+S^2\cos q$ in agreement with the low temperature
expression obtained from the exact result and hence no high momentum peak
arises in this limit.}
\end{thebibliography}
\end{document}